\documentstyle[aps,epsf,floats,preprint]{revtex}
\tighten
\begin{document}
\title{Geometric entropy, area, and strong subadditivity}
\author{H. Casini}
\address{The Abdus Salam International Centre for Theoretical Physics, \\
I-34100 Trieste, Italy \\
e-mail: casinih@ictp.trieste.it}
\maketitle

\begin{abstract}
The trace over the degrees of freedom located in a subset of the space
 transforms the vacuum state into a mixed density matrix with non zero entropy.
 This geometric entropy is believed to be deeply related to the entropy of
 black holes. Indeed, previous calculations in the context of quantum field theory,
 where the result is actually ultraviolet divergent, have shown that the geometric entropy
 is proportional to the area for a very special type of subsets.
 In this work we show that the area law follows in general from
 simple considerations based on quantum mechanics and relativity.
 An essential ingredient in our approach is the strong subadditive property
 of the quantum mechanical entropy.
\end{abstract}

\section{Introduction}
A chain of  arguments using thermodynamics, classical general relativity and
 quantum field theory in curved spaces shows that the black hole has an associated 
entropy proportional to the
 horizon area. This is arguably one of the most important clues to the understanding 
of the role of gravity at the quantum level. However, these calculations
give no precise identification of
the degrees of freedom responsible for the entropy, nor an explanation of its relation 
with area. There are several approaches in this regard, involving either
 specific models of quantum gravity or effective ideas \cite{wa}.

 A very appealing candidate in this last sense is given by the entropy of entanglement \cite{th,ent}. 
On the black hole spacetime we can write the Hilbert space of   
a quantum field as a tensor product of two Hilbert spaces, which correspond  
to the asymptotic data at infinity and at the horizon respectively. 
The quantum state relevant
 for asymptotic observers is obtained by tracing the complete quantum state density matrix 
 $\vert \Psi\rangle \langle\Psi\vert$  
 over the invisible degrees of freedom on the horizon. But $\vert \Psi\rangle$
 contains correlations between the field modes inside and outside the black hole, 
or, in other words, it is entangled with respect to the above tensor product. 
Thus, the partial trace leads to a mixed density matrix with non zero entropy.

For large black holes the near horizon geometry is approximately flat. 
This suggests that the relevant entropy should also occur in flat space \cite{w}.  
 To mimic the black hole in Minkowski space take a spatial hyperplane and a subset $X$ on it. 
This divides the space in two, the interior and
 exterior regions with respect to $X$. The trace of the vacuum state over the interior (or exterior) 
degrees of freedom is a mixed density matrix. Its entropy, being a function of the subset $X$, 
 was called geometric entropy \cite{w}. This new flat space quantity,
 suggested by black hole physics, is interesting in its  
own right, and provides a simpler context to contrast ideas.   

Several authors have done explicit calculations
 of the geometric entropy in quantum field theory (QFT). 
 The methods ranged from numerical work \cite{ent,sred}, to analytical calculations based on
the replica trick, the heat kernel, and conformal symmetry in two dimensional 
spacetime \cite{w,w1,sol,ka1,con,mh}. In all cases, the reassuring result is an entropy  
proportional to the area.
 However, most of the work has focused on free fields and 
 a very restricted class of subsets. The application of the analytical methods was constrained to
  the infinite half space since they use in an essential way the 
fact that density matrix corresponding to the Rindler wedge is a thermal one \cite{mh}.
 The numerical methods encounter practical problems for non spherical subsets (see however the recent 
 work on annular and flower-like geometries \cite{b11,bo}).    
 Also, the physical principles behind the proportionality between area and entropy remain obscure 
beyond the explicit calculations, that involve dealing with non renormalizable divergences.   

In this work we show that for a wide class of subsets the area law for the geometric entropy 
follows from very general and model independent ideas based on quantum mechanics and relativity.  

We assume, as it has been done explicitly or implicitly in all the previous work on the subject, 
 that it is possible to divide a Cauchy surface into pieces and trace the vacuum state 
 over the degrees of freedom in one of them to obtain a density matrix. This is the single
 basic condition allowing to define a geometric entropy. However, this is 
 a delicate point, related to the divergences appearing in the QFT calculations. We will only 
say a bit more about it in Section III. 

The causal evolution is unitary. Accordingly, the geometric entropy 
corresponding to two space-like subsets coincide if they give place to the same causal 
domain of dependence, or, what is the same, they can be extended to a global Cauchy surface by
 union with the same space-like subset. We also make use of the Poincare symmetry of the vacuum state.  
 
It is essential to the arguments below the mathematical fact that the quantum mechanical entropy is
 positive and strongly subadditive.
The strong subadditive property was introduced in \cite{rob,intro}, in the context of the investigations
around an old problem in statistical mechanics, 
namely, the proof of the existence of the mean entropy (the limit of entropy over volume for large sets).
 Here we apply these ideas to the relativistic situation. 
Note that for a general quantum statistical mechanical system the entropy in a given volume
 has to be defined in the same way as the geometric entropy is defined 
 above, that is, by tracing over degrees of freedom in the appropriate domain.
In this sense, the geometric entropy is just the name for the more familiar concept of the
 entropy contained in a volume in the special case where the state is the relativistic vacuum.

The cited investigations on the Euclidean symmetric situation where
 mainly focused on the extensive properties of the entropy on large domains. Here the mean entropy 
is zero. Basically, this is so because we can construct
Cauchy surfaces with arbitrarily small volume by approaching the null surfaces, and use
the boost symmetries and the subadittive property to put an upper bound to the entropy 
which grows like the area. Moreover, the strong subadditivity allows to completely determine
 the functional form of the geometric entropy. Curiously, 
a very relevant piece of the information it provides, that gives the clue to show 
 that the geometric entropy 
is not only bounded by the area but linearly growing with it, 
  shows up already when combined only with translation invariance. 
However, the crucial limit for the geometric entropy is not the one of infinite volume but
 that of the very flat type of sets. 

Summarizing, we show that the combination of strong subadditivity, Poincare symmetry and causality constrains the
geometric entropy to be proportional to the bounding area, plus a constant term.

The work is organized as follows. In Section II we review some properties of the quantum entropy.
In Section III we state the problem of the geometric entropy in Euclidean and relativistic spaces.
 In Section IV we review some work on the entropy for Euclidean symmetric states and
develop some new results on this subject, related to the entropy of very
flat sets. 
The main result of the paper is the theorem in Section V which states that geometric 
entropy is proportional to the area.  
Finally we present the conclusions and discuss some open perspectives. 

\section{Properties of the quantum entropy}

In this Section we review some inequalities satisfied by the quantum
entropy. The proofs of the statements,
the original references and further details can be found in the reviews \cite{Wehrl,l}.

In quantum mechanics a physical state is described
by a density matrix $\rho$,
which is a self-adjoint and positive operator with unit trace, $Tr \rho=1$,
defined on a Hilbert space ${\cal H}$. The entropy $S$ is a function of the quantum
state given by the expression 
\begin{equation}
S=-Tr \rho \log \rho \,.  \label{entro}
\end{equation}
The quantum entropy is always positive, $S\ge 0$, being zero if and only if
the state is pure, that is, the density matrix is a one dimensional
projector $\rho=\vert \Psi\rangle \langle\Psi\vert$. The entropy can be
infinite for infinite dimensional Hilbert spaces, while on a space with
finite dimension $d$ it is bounded above by $\log d$. The function $S$ satisfies 
\begin{equation}
\lambda_1 S(\rho_1)+\lambda_2
S(\rho_2)-\lambda_1 \log (\lambda_1) -\lambda_2 \log (\lambda_2) 
\ge S(\lambda_1 \rho_1+\lambda_2 \rho_2)\ge \lambda_1 S(\rho_1)+\lambda_2
S(\rho_2) \, ,\label{sll}
\end{equation}
where $\lambda_1$ and $\lambda_2$ are any positive numbers such that $
\lambda_1+\lambda_2=1$. The second inequality means that the effect of mixing always increases the
entropy. Moreover, any state can be decomposed as a mixing of pure states.

In the following we will be mainly interested in the properties of the
entropy related to spaces constructed out of the tensor product of two or
more Hilbert spaces. Let the Hilbert space ${\cal H}={\cal H}_1 \bigotimes {\cal H}_2$ and $
\rho=\rho_{i_1 i_2 \, ,i_1^{\prime} i_2^{\prime}}$ be the density matrix on $
{\cal H}$. By tracing over ${\cal H}_1$ we can form the density matrix $\rho_2\equiv
(\rho_2)_{i_2 i_2^{\prime}}=\sum_{i_1} \rho_{i_1 i_2 \, ,i_1 i_2^{\prime}}$
on ${\cal H}_2$. The density matrix $\rho_1$ is defined similarly by tracing over $
{\cal H}_2$.

For a pure density matrix $\rho$ the following duality property for the
entropy of the subsystems holds 
\begin{equation}
S(\rho_1)=S(\rho_2) \,.  \label{dual}
\end{equation}
This follows from the fact that for $\rho=\vert \Psi\rangle \langle\Psi\vert$
we have the equality $Tr (\rho_1^n)= Tr (\rho_2^n)$ for any power $n$, and
consequently the non zero eigenvalues and their multiplicities coincide for $
\rho_1$ and $\rho_2$.

It is easy to give examples where $\rho$ is pure, so that $S(\rho)=0$, while 
$\rho_1$ is in a mixed state with $S(\rho_1)> 0$. Thus, the quantum entropy
can not be generically increasing (monotonicity) with the size (the number of
degrees of freedom) of the subsystem (nor with the size of the subsystem
that is traced over). This is somehow against the intuition coming from the
extensive entropy of gases.

A natural question is what type of density matrices $\rho_1$ can arise from
tracing a pure density matrix $\rho$ over a subsystem. The answer is that
any density matrix can be obtained in this way. A density matrix $\rho_1$ on
a space ${\cal H}$ can always be realized as the partial trace of a pure density
matrix $\rho$ on ${\cal H}\bigotimes {\cal H}$.

The entropy for a system composed by two independent subsystems is the
 sum of the subsystem entropies. This is at the physical basis of
the explicit form of $S$ given by equation (\ref{entro}). The density
matrix for the composed system is given by the tensor product $
\rho=\rho_1\bigotimes \rho_2$ in this case, and we have the equation 
\begin{equation}
S(\rho)=S(\rho_1)+S(\rho_2)\,.
\end{equation}

In the general case where $\rho$ does not factorize and the subsystems can
not be considered as statistically independent, this equation can be extended
to an inequality expressing the subadditive property of the entropy 
\begin{equation}
S(\rho)\le S(\rho_1)+S(\rho_2)\,.   \label{saa}
\end{equation}

In fact the entropy satisfies a stronger inequality. To introduce it, it is
convenient to consider a more general case in which the Hilbert space of the
system is a tensor product of an arbitrary number of factors $
{\cal H}=\bigotimes_{i\in I} {\cal H}_i$, where $I$ is the set of indices labeling
the different subsystem Hilbert spaces ${\cal H}_i$. Let $A$ be any subset of the
set of indices $I$. Define the corresponding reduced density matrix on $
\bigotimes_{i\in A} {\cal H}_i$, by tracing over all ${\cal H}_i$ with $i\notin A 
$, 
\begin{equation}
\rho_A=Tr_{\bigotimes_{i\in -A}{\cal H}_i} \rho\,.
\end{equation}
Call $S(A)\equiv S(\rho_A)$ the corresponding entropy. With this notation
 the subadditive property (\ref{saa}) 
writes
\begin{equation}
S(A)+S(B)\ge S(A\cup B)\,  \label{sub}
\end{equation}
for any pair of subsets $A$ and $B$ of $I$, while the
duality relation for a pure total density matrix $\rho$ is simply 
\begin{equation}
S(A)=S(-A)\,.  \label{menos}
\end{equation}
Here $-A$ is the set complementary to $A$ in $I$. As mentioned, $I$
can be artificially enlarged to a bigger set $I^{\prime}$,
and the total density matrix can be taken
 pure in the enlarged space, without modifying the value of $S(A)$ for
  $A\subseteq I$. In this case (\ref{menos}) holds in the bigger space
$I^{\prime}$.

Applying subadditivity
 to the complements $-A$ and $-B$
 we have $S(-A)+S(-B)\ge S(-(A\cap B))$,
 since the complement operation interchanges union and intersection.
Thus, purifying the total density matrix the inequality (\ref{saa})
 implies a different relation,
namely
\begin{equation}
S(A)+S(B)\ge S(A\cap B).       \label{ff}
\end{equation}
The strong subadditivity (SSA) generalizes the relations (\ref{sub})
and (\ref{ff}) 
in a form that is self-dual under taking complements in the pure case.
It writes
\begin{equation}
S(A)+S(B)\ge S(A\cup B)+S(A\cap B)\,.  \label{ss}
\end{equation}

Using the trick of working with an artificially purified density matrix
 the SSA applied to $A$ and $-B$ leads to a different self-dual relation.
 This is
\begin{equation}
S(A)+S(B)\ge S(A-B)+S(B-A)\,,  \label{ultima}
\end{equation}
where $A-B$ means $A\cap (-B)$. The inequalities (\ref{ss}) and
(\ref{ultima}) are equivalent in the pure case but they are both valid
in general. In lack of a better name, and since it is closely related to
SSA, we often call SSB the inequality (\ref{ultima}).

Finally, we also mention that for two non intersecting sets $A$ and $B$
 the SSA and SSB relations lead to the following triangle inequalities
  for the entropies
\begin{equation}
\vert S(A) -S(B) \vert \le S(A\cup B) \le S(A)+S(B)\,.  \label{tri}
\end{equation}
The second one is of course subadditivity. The inequalities (\ref{tri}) are 
 useful to prove some continuity properties of $S$. For example, if the
 entropy of $B$ is very small one has that $S(A)$ and $S(A\cup B)$ are very
 near to each other.

We mentioned that $S$ is not monotonic in general. By monotonicity here
 we mean the property $S(B)\le S(A)$ for any $B\subseteq A$.
 However, the inequality (\ref{ultima}), or the first one from
 (\ref{tri}), can be
thought as partial compensations for the lack of monotonicity.

Inequalities provide less concrete information and are more difficult to
handle than equations. This is perhaps the reason why the relations (\ref{ss}) and
(\ref{ultima})  are
relatively poorly known and have found only few applications. However, in the presence
of symmetries these can be very useful as will become apparent in the
 following.

\section{Strong subadditivity and continuum quantum systems}

Let us consider a continuous quantum system in ${\bf R}^{d}$ described in terms of
a Fock space. The single particle states are given by elements of $
{\cal H}_{1}({\bf R}^{d})={\cal L}^{2}({\bf R}^{d})$, the Hilbert space of square integrable functions
in ${\bf R}^{d}$, while the $n$ particle space is the product of $n$ copies of $
{\cal H}_{1}({\bf R}^{d})$, ${\cal H}_{n}({\bf R}^{d})={\cal H}_{1}({\bf R}^{d})\bigotimes ...\bigotimes
{\cal H}_{1}({\bf R}^{d})$, where the tensor product is understood as symmetrized or
antisymmetrized according to the particles being bosons or fermions. The Fock
space is then given by the direct sum ${\cal H}=\bigoplus_{0}^{\infty }{\cal H}_{n}$. Similarly, we can
define the single particle states in a given volume $V$ in ${\bf R}^{d}$ as $
{\cal H}_{1}(V)={\cal L}^{2}(V)$, and its corresponding Fock space ${\cal H}(V)=\bigoplus_{0}^{
\infty }{\cal H}_{n}(V)$. If the intersection of two sets $V_{1}$ and $V_{2}$
has zero measure
 we have  
\begin{equation}
{\cal H}({V_{1}\cup V_{2}})={\cal H}(V_{1})\bigotimes {\cal H}(V_{2})\,.  \label{a}
\end{equation}
Now, let the density matrix of the system be $\rho $. We can define a
density matrix $\rho _{V}$ for a volume $V$ by tracing $\rho $
over ${\cal H}(-V)$. Its  entropy is $S(V)=-Tr\rho _{V}\log (\rho
_{V})$. Thus, the density matrices corresponding to the subsets satisfy the
compatibility condition 
\begin{equation}
\rho _{V}=Tr_{{\cal H}(V^{\prime })}\rho _{V\cup V^{\prime }}\,,  \label{b}
\end{equation}
where $V^{\prime }$ is any set that has measure zero intersection with $V$.
A continuous quantum system is defined here as a set
of Hilbert spaces and density matrices corresponding to subsets of ${\bf R}^d$
 satisfying the conditions (\ref{a}) and (\ref{b}). The strong subadditive
 inequalities 
then take a geometric form given by (\ref{ss}) and (\ref{ultima}),
where now $A$ and $B$
represent subsets of ${\bf R}^d$ \cite{Wehrl}.

The problem of building a continuous quantum system can be presented in a slightly 
different way where the complete system is constructed out of the finite 
 volume subsystems. These 
 may be defined for example by giving a Hamiltonian and boundary conditions, and taking a finite 
volume density matrix $\rho_V$. Then, to glue all elements into  
a unique system we have to impose (\ref{a}) and (\ref{b}) on the local Hilbert spaces and states.  

When the total state $\rho$ is invariant under some transformation ${\cal U}$ in ${\bf R}^d$
 the entropy  $S({\cal U}(A))$ is equal to $S(A)$. This symmetry combined with the properties of 
the entropy can strongly constrain the function $S$. We say that a positive function $S$ on 
 subsets of ${\bf R}^d$ is an Euclidean entropy
function if it is invariant under rotations and translations, and satisfies SSA and SSB.    

\subsection*{Relativistic setting}

In the relativistic case a pair of time-like related subsets contains
 degrees of freedom that are not mutually independent. A scheme as the previous one for an
Euclidean system, where (\ref{a}) and (\ref{b}) hold, must correspond to every
Cauchy surface in spacetime.

All previous work on geometric entropy assumes that
we can divide a Cauchy surface for the spacetime into pieces
 and trace the total state over the degrees of freedom on one of them to obtain a density matrix.
 This assumption is also made here, since it is fundamental to the very definition of a geometric entropy.

However, this is at the root of the divergences encountered in
 quantum field theory calculations. The complete localization of states in a given region is forbidden by standard
axioms in QFT (see \cite{ha} for a clear discussion on this point). The covariant regularizations that deal with
the infinites arising in perturbation theory can not in general avoid the divergence in the entropy,
which shows up at the level of free fields.
Also, this is not softened by supersymmetry.
Thus, as these divergences suggest, there are two types of enigma related to the black hole entropy
formula. The first is given by the proportionality between area and entropy, and the second one is that the
 entropy is surprisingly finite. Here we do not deal with this last and probably deeper problem.
It is possible that a finite entropy would require quantum gravity
input (see also a different idea in \cite{be}).
 However, the existence of a good covariant
definition for a finite geometric
entropy in the context of QFT can not be discarded. But, in that case, the role of the set boundaries
 can not be taken as purely passive. In other words, to define a bounded system Hilbert space,
boundary conditions are needed. The precise compatibility conditions that these bounded systems should
 have in order to be used for computing a geometric entropy are given by (\ref{a}) and (\ref{b}).
This will be enough characterization for our purposes.
Any non ambiguous definition for a localized entropy beyond the classical limit should suffer from
 the same problem. In particular, this is the case for the various entropy bounds present in the literature,
which are suggested by ideas related to the physics of black holes \cite{b1}.
In the semiclassical level they should take into account the geometric entropy of quantum origin, what
 can make them much more constraining and interesting \cite{ss}.

\begin{figure}
\centering
\leavevmode
\epsfxsize=12cm
\epsfbox{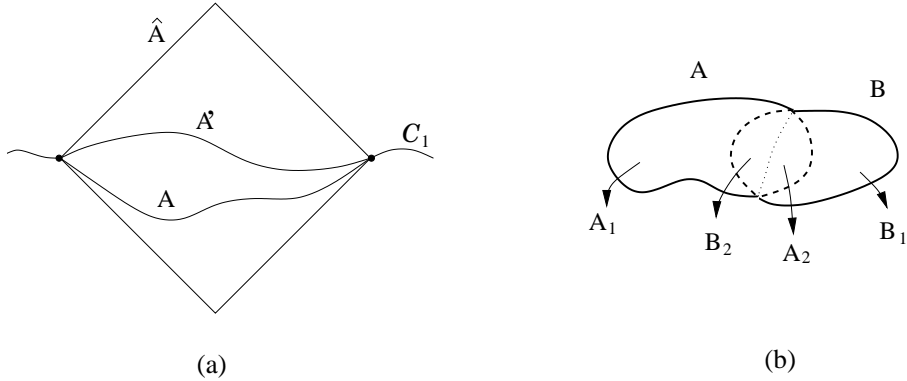}
\bigskip
\caption{(a)- A causally closed set $\hat{A}$ in $1+1$ dimensions. It is the causal development
of the spatial surfaces $A$ and $A^\prime$. We say that $A$ and $A^\prime$ are Cauchy surfaces for $\hat{A}$.
 Both of these sets can be continued to a Cauchy surface for the Minkowski space using the same spatial set
${\cal C}_1$. The marked points on the left and right corners of the diamond shaped set $\hat{A}$ represent
 the spatial corner of $\hat{A}$. (b)- Two commuting causally closed sets in $2+1$ represented here by their
Cauchy surfaces representatives $A$ and $B$. These surfaces intersect in the dotted line in this example,
dividing $A$ into $A_1$ and $A_2$ and $B$ into $B_1$ and $B_2$.
 The commutation
 imposes that $A_2\subseteq \hat{B}$ and $B_2\subseteq \hat{A}$. Note that the spatial corners of $\hat{A}$
and $\hat{B}$, which are the boundaries of $A$ and $B$ respectively, are spatial to each other.
 The causally closed sets $\hat{A}\vee \hat{B}$ and $\hat{A}\wedge \hat{B}$ that enter into
the SSA relation are generated by $A_1\cup B_1$ and $A_2\cup B_2$ respectively.}
\label{rela1}
\end{figure}

Therefore, here we assign a density matrix and an entropy to any subset $A$ of a Cauchy surface ${\cal C}$.
 In the following we restrict the analysis to the $A$ with compact closure.
 The conditions (\ref{a}) and (\ref{b}) are assumed to hold among the subsets of any such surface ${\cal C}$.

The causal development $\hat{A}$ of $A$ is the set where the solution of a wave like equation
is uniquely determined by the initial data on $A$. More generally, for any set $O$ not necessarily space-like,
 the causal development is the set of points $x$ such that every inextensible time-like curve through $x$ 
 cuts $O$. 
We often say, abusing of the terminology, that a space-like set $A$ is a Cauchy surface for $\hat{A}$.
 Consider two subsets $A$ and $A^\prime$ of 
 different Cauchy surfaces ${\cal C}$ and ${\cal C}^\prime$, but having the same 
 causal development $\hat{A}=\hat{A}^\prime$ (see fig.(\ref{rela1})a). We can extend $A$ and $A^\prime$ 
with the same set to form a global Cauchy surface, that is, we can choose 
${\cal C}_1={\cal C}-A = {\cal C}^\prime-A^\prime$. 
 The unitarity of the causal evolution implies that we have 
\begin{equation}
S(A)=S(A^\prime)\equiv S(\hat{A})\,.
\end{equation}
 This means that there is no loss nor gain of information in passing from $A$ to $A^\prime$.  
In the Heisenberg picture both $\rho(A)$ and $\rho(A^\prime)$ are obtained by tracing 
the total state over the same degrees of freedom in ${\cal C}_1={\cal C}-A = {\cal C}^\prime-A^\prime$.

 Then, $\hat{A}$ can be seen as an equivalence class of pieces of Cauchy surfaces with the same 
causal development, density matrix and entropy. Note that all Cauchy surfaces for $\hat{A}$ share 
the same boundary, the spatial corner of
$\hat{A}$ (the subset of the closure of $\hat{A}$ that is not time-like related to any point in $A$, see the 
figure (\ref{rela1})). 
The sets of the type $\hat{A}$ 
 are called here causally closed, and they constitute the natural domain of $S$ in the
  relativistic case.    
 We often make no distinction between the set $\hat{A}$ and an element 
 $A$ on the equivalence class it represents, and denote them with the same capital letter.

 Now we identify the conditions allowing to apply the strong subadditive property for
 any two given causally closed sets $\hat{A}$ and $\hat{B}$. As a consequence of eqs. (\ref{a}) and (\ref{b}) the
 SSA and SSB inequalities in the form (\ref{ss}) and (\ref{ultima}) take place for the entropies of two 
subsets $A$ and $B$ of the same global Cauchy surface ${\cal C}$. 
In that case the sets $A \cap B$ and  $A \cup B$ appearing 
 in the smaller side of the strong subadditive relation are Cauchy surfaces for 
$\hat{A}\cap \hat{B}$ and the 
the causal development of $\hat{A}\cup\hat{B}$ respectively.
However, it is possible to apply SSA and SSB even in 
some cases where $A$ and $B$ do not belong to the same ${\cal C}$. As the entropies really depend
 on the equivalence classes $\hat{A}$ and $\hat{B}$, this happens if there are  
different Cauchy surfaces $A^\prime$ and $B^\prime$ for $\hat{A}$ and $\hat{B}$  
 that belong to the same global Cauchy surface ${\cal C^\prime}$. 
 We say that two causally closed sets $\hat{A}$ and $\hat{B}$ (or any pair of Cauchy surfaces for them) 
commute if there is at least a pair of 
the respective Cauchy surface representatives that belong to the same global Cauchy surface 
(see fig.(\ref{rela1})b). 
 It can be seen that an equivalent condition for the  commutativity of $\hat{A}$ and $\hat{B}$ is that their  
spatial corners are spatial to each other. In particular, if $\hat{A}$ 
includes $\hat{B}$ or they are spatially separated, they commute.  
The inequalities SSA or SSB can be applied only for the entropies of commuting elements. For these, 
 the right hand side of (\ref{ss}) and (\ref{ultima}) has to be calculated with any pair of 
  Cauchy surface representatives belonging to the same global Cauchy surface.

There is a more elegant way to define the concept of causally closed sets 
and commutativity that has interesting connections with quantum mechanics \cite{h}.
Write $x\sim y$ for two points that are time-like related. Given any subset $O$ of the spacetime
 define its causal opposite $O^\perp$ as the subset of points which are spatial to $O$, 
$O^\perp=\{ x/x \not \sim y, x\neq y, \text{for every}\, y\in O\}$. The
 causally closed sets are just the subsets $O$ which coincide with their double opposite,
 $O^{\perp\perp}=O$~\footnote{This equivalence holds for bounded sets. 
The causally closed sets $O=O^{\perp\perp}$ are exactly the domain of dependence of achronal sets bounded in time
 \cite{cw1}.}. Curiously, this sets form an orthomodular lattice (also called quantum logic).
 This structure is also shared by the physical propositions in quantum mechanics, or
equivalently, the orthogonal projectors on the Hilbert space. In any orthomodular lattice
 there are three operations, the opposite, the meet $A\wedge B$ and the join $A\vee B$ 
of two elements, and an order relation $\subseteq$. In the present context they are given 
respectively by the above opposite, the intersection $A\cap B$,
 the double opposite of the union $(A\cup B)^{\perp\perp}$, and the set  inclusion 
relation $\subseteq$. There is a definition of commutativity of two elements
 for any orthomodular lattice that coincides with the usual commutativity of operators when 
applied to projectors in the Hilbert space. Two elements $A$ and $B$ commute when
 \begin{equation}
A=(A\wedge B)\vee (A\wedge B^\perp)\,.
\end{equation} 
This definition coincides with the one given above for the commutativity of 
causally closed sets. Thus, the 
compatibility requirement that two causally closed sets have Cauchy surfaces which can be extended 
 to a common global one can be put in the same mathematical terms as the 
compatibility condition for two projection operators to be simultaneously measurable. 
With this notation, the SSA and SSB relations for two commuting sets $A$ and $B$ are
\begin{eqnarray}
S(A)+S(B) & \ge & S(A\vee B)+S(A\wedge B)\,,\\
S(A)+S(B) & \ge & S(A\wedge B^\perp)+S(B\wedge A^\perp)\,,
\end{eqnarray}  
 respectively.
  
A relativistic entropy function is defined as a positive\footnote{There was an indication  
of a negative geometric entropy when gauge fields 
are present \cite{w1}. This troubling result  was finally corrected \cite{ka1,mh}.}
 function on the causally closed sets in Minkowski space, symmetric under Poincare transformations,
 and satisfying SSA and SSB for commuting 
pairs of subsets. The use of the relation SSB
 in the proofs below may be avoided, but it considerably simplify the demonstrations. 
 
Thus, the problem of finding the functional dependence of the geometric entropy 
takes a precise mathematical form, being equivalent to find the most 
general relativistic entropy function. We note that 
 a linear combination of solutions with positive
 coefficients is also a solution. 

A similar problem can be stated in any space-time, but in absence of the Poincare
symmetry the number of solutions can be very big. For example, the flux of any 
future directed conserved current through a Cauchy surface for a causally
 closed set $A$ is always a solution of SSA and SSB since it satisfies the  
 strong subadditive equation (rather than inequality). Of course 
 this solution is not Poincare invariant in Minkowski space. 
 Symmetries with orbits that are exclusively time-like (or space-like) do not 
impede this type of extensive solutions.  

In this paper we are only using causally closed sets with compact closure, 
and, as in the Euclidean case, make no distinction between sets differing
by zero measure subsets. The game to compare the entropies on different sets
using strong subadditivity is more subtle than in an ordinary additive
measure theory. Because of that we further restrict the class of subsets.
To see what are the difficulties about, consider first the polyhedra on the Euclidean space ${\bf R}^d$.
The intersections and unions of polyhedra are again polyhedra. Thus, the relations
 SSA and SSB can not be used to obtain information for the entropy of sets
 with curved borders from the entropy on polyhedra. In order to focus on the
 most restrictive relations coming from the entropy properties and the symmetries,
 in the following
we consider the domain of the Euclidean entropy functions to be the class
of all polyhedra. Note, however, that these can have an arbitrarily large number of faces
 as small as we want, and, in this sense, we can approach any set by polyhedra.
To connect the results about polyhedra with other kind
 of sets some continuity condition is needed, and we do not deal with this problem here.
 In the relativistic $d+1$ dimensional case we consider the class of relativistic polyhedra, that is,
 spatial $d$ dimensional sets (or their corresponding $d+1$ dimensional causally closed sets)
formed by union of a finite number of polyhedra on spatial hyperplanes. These also form a
closed class under union and intersection when taken between commuting sets.

A note on the terminology. We call $\text{vol}(X)$ the volume of a $d$ dimensional
 Euclidean polyhedron $X$,
 and $\text{area}(X)$ to the area of the $d-1$ dimensional boundary of $X$. For a relativistic
  $d+1$ dimensional polyhedron $X$ we call $\text{area}(X)$ the area of its spatial
corner, which is the boundary of any Cauchy surface for $X$. A relativistic entropy
 function $S$ on $d+1$ dimensions when restricted to polyhedra on a unique hyperplane gives place to an Euclidean
 entropy function on $d$ dimensions. This latter is independent of the chosen hyperplane.
When it is possible, we identify these functions, and call them with the same letter.

\section{Entropy in an Euclidean symmetric system}
\subsection{The one dimensional case}

Consider a system on the real line and a translational invariant
density matrix $\rho $. The connected compact sets on $
{\bf R}^1$ are intervals, determined up to a translation by their length $x$. On
these intervals
the entropy can be written as a function on a real variable $S(x)$.
The results we obtain here about $S(x)$ can be found for example
in the ref. \cite{Wehrl}.

Take two intervals of length $(x+y)/2$, with $x$ and $y$ positive and $
y>x$, and arrange them so that their intersection has length $x$ and
consequently their union has length $y$. Applying strong subadditivity we
have
\begin{equation}
S\left(\frac{x+y}{2}\right)\ge \frac{1}{2}S(x)+\frac{1}{2}S(y)\,.  \label{weak}
\end{equation}
Suppose now that $S(x)$ is infinite. Then it follows from (\ref{weak}) that
$S(z)$ must be infinite for $(x/2)\le z\le \infty $. Repeatedly using this
argument we have that the function $S(x)$ is either finite for all $x$ or
always infinite.

Similar arguments can be used to show that all the entropy functions
 considered in the paper are either infinite or always finite. We are not
 coming back to this point and assume a finite $S$ in the following.

\begin{figure}
\centering
\leavevmode
\epsfysize=4cm
\epsfbox{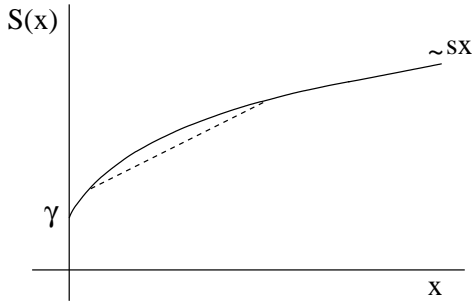}
\bigskip
\caption{A non decreasing
concave function $S(x)$. Two points on the curve $S(x)$ determine
 a segment that lies below the function graph. The limit of $S(x)$ for $x\rightarrow 0$
 is $\gamma$, and the curve slope approaches to $s$ as $x$ goes to infinity (however,
 $S(x)$ does not necessarily approach any straight line asymptotically).}
\label{conca}
\end{figure}

The relation (\ref{weak}) for the function $S(x)$ is called weak concavity,
while the concavity is the property
\begin{equation}
S(\lambda x+(1-\lambda )y)\ge \lambda S(x)+(1-\lambda )S(y)\,,
\label{concave}
\end{equation}
for any $0\le \lambda \le 1$. A function is concave when the segment joining
any two points in its graph lies below the function curve (see the fig.(\ref{conca})).
Weak concavity is in
fact very near concavity, since applying (\ref{weak}) repeatedly it is easy
to obtain eq.(\ref{concave}) for any rational $\lambda $ with a power of $2$
as a denominator. For example, we have $S(3/4x+1/4y)\ge \frac{1}{2}S(x)+
\frac{1}{2}S(1/2x+1/2y)\ge \frac{3}{4}S(x)+\frac{1}{4}S(y)$. To show that $
S(x)$ is actually concave suppose that $x<z<y$ and, violating concavity,
\begin{equation}
S(z)=\frac{y-z}{y-x}S(x)+\frac{z-x}{y-x}S(y)-\delta \,,  \label{paro}
\end{equation}
with $\delta >0$. Let $w=\lambda x+(1-\lambda )y$ with $\lambda \in [0,1]$ a
rational with a power of $2$ as a denominator. We can take $w=z+\epsilon $
with $\epsilon $ a positive number as small as we want. Positioning two
intervals of length $z$ such that their union has length $
z+\epsilon =w$, and their intersection has length $z-\epsilon $, we have
from SSA and (\ref{paro})
\begin{equation}
S(z-\epsilon )\le S(z)-\delta +{\cal O}(\epsilon )\,.  \label{poro}
\end{equation}
Thus, as near to $z$ as we want there are points in which, let say, $S<S(z)-
\frac{1}{2}\delta $. By repeating this argument for a sufficient number of
times we could arrive at the result that there would be points as near to $z$
as we want that have negative $S$. As the entropy is positive, equation
(\ref{paro}) must be wrong, and the function $S$ is concave, equation
(\ref{concave}) holds for any $\lambda \in [0,1]$. The converse is also true,
concavity implies strong subadditivity for intervals.

Suppose that for some $y>x$ it is $S(y)< S(x)$. Then the function $
S$ is decreasing for any $z>y$, because a consequence of concavity is that the
slope of the function $S$ always decreases. This would lead to $S(z)<0$ for
high enough $z$. To avoid it the function $S$ must be non decreasing
everywhere. The monotonicity of the entropy appears here as a
 consequence of translation symmetry.
However,  order by inclusion is not
warranted for the entropies of sets formed by several disjoint
intervals, or connected sets in more dimensions.
The concavity also imply that $S(x)$
 is continuous.

Being nondecreasing and positive $S(x)$ has a well defined limit as
$x$ goes to zero. We call $\gamma$ this limit. Given $x<y$ we have
 from concavity that
\begin{equation}
S(x)\geq \gamma+x\frac{S(y)-\gamma}{y}\,.
\end{equation}
Since $\gamma\ge 0$ it follows that $S(x)/x$ is a non increasing function.
This leads to the existence of the limit $\lim_{x\rightarrow \infty
}S(x)/x=s $, with $s\ge 0$. Therefore there exist a well defined notion of mean entropy in
the system (otherwise $S(x)/x$ could increase to infinity or oscillate
indefinitely). For big enough sets the entropy is approximately extensive, 
at least if $s\neq 0$. 

\subsection{The limit for the entropy of small one dimensional sets}

\begin{figure}
\centering
\leavevmode
\epsfxsize=13cm
\epsfbox{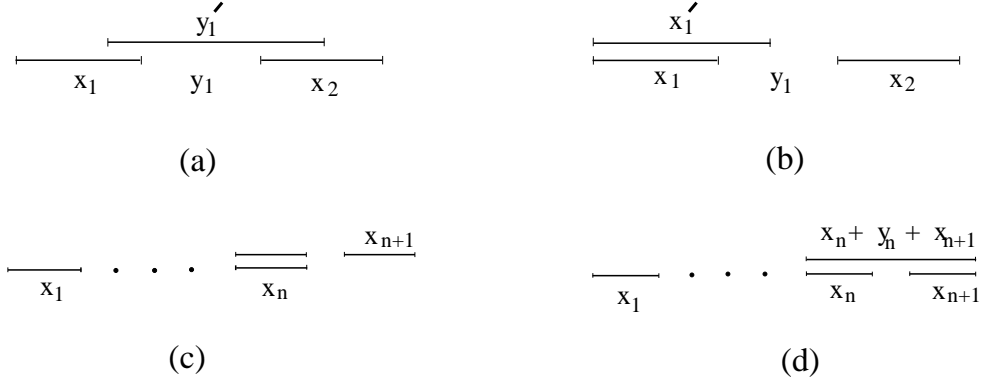}
\bigskip
\caption{Geometric constructions used to prove the formula (\ref{limite})
 for the entropies of one dimensional sets in the limit of small size.}
\label{unid}
\end{figure}

The measure zero sets have zero entropy. However, sequences of sets with
measures converging to zero can have non zero entropy limit. The limit of $
S(x)$ when $x$ goes to zero is in general a positive number $\gamma$.
Consider now the case of a set formed by two intervals of lengths $x_1$ and $
x_2$ separated by a distance $y_1$. We are interested in the limit $
l=x_1+y_1+x_2\rightarrow 0$. If $\epsilon$ is a small positive number we can
choose $l$ such that $S(l)\le \gamma+\epsilon$. Placing an interval of length $
y_1^\prime\le l$, $y_1^\prime\ge y_1$ as shown in fig.(\ref{unid})a it follows from
SSA that 
\begin{equation}
S(x_1,y_1,x_2)+S(y_1^\prime) \ge S(x_1^\prime ,y_1,x_2^\prime )+S(l) \,,
\end{equation}
 with $x_1^\prime\le x_1$, $x_2^\prime\le x_2$, $x_1^\prime +x_2^\prime +y_1=y_1^\prime$. 
Taking into account the monotonicity of $S(x)$ we have 
\begin{equation}
S(x_1,y_1,x_2)\ge S(x_1^\prime ,y_1,x_2^\prime ) \,,  \label{auno}
\end{equation}
for any $x_1^\prime \le x_1$ and $x_2^\prime\le x_2$. Similarly, from the same
construction of fig.(\ref{unid})a but now using SSB we have 
\begin{equation}
S(x_1,y_1,x_2) \ge S(x_1^\prime ,y_1^\prime ,x_2^\prime ) -\epsilon\,,
\label{ados}
\end{equation}
for any $y_1^\prime\ge y_1$, $x_1^\prime \le x_1$, $x_2^\prime\le x_2$, and $
x_1+y_1+x_2=x_1^\prime +y_1^\prime +x_2^\prime$. Thus, reducing the length of the 
two intervals reduces the entropy (neglecting contributions of order $\epsilon$).
The converse is given by the construction of figure (\ref{unid})b. Applying strong
subadditivity we get 
\begin{equation}
S(x_1, y_1, x_2)\ge S(x_1^\prime ,y_1^\prime , x_2 )-\epsilon \,,
\label{atres}
\end{equation}
for any $x_1^\prime \ge x_1$, $y_1^\prime \le y_1$, and $x_1+y_1=x_1^\prime
+y_1^\prime$.

From (\ref{auno}), (\ref{ados}) and (\ref{atres}) we see that $
S(x_1,y_1,x_2)$ converge to a limit $\delta$ when $x_1+y_1+x_2$ go to zero,
and this limit is independent of the way in which the interval lengths $x_1$
and $x_2$ go to zero.

Now consider the entropy $S(x_1,y_1,...,y_{m-1},x_m)$ of
a set formed by $m$ disjoint intervals of lengths $x_1,...,x_m$ separated by
distances $y_1,...,y_{m-1}$. We analyze the entropy in the limit when $
l=x_1+y_1+...+x_m$ go to zero and prove by induction that this depends only
on the number of connected components $m$, and is given by 
\begin{equation}
\lim_{l\rightarrow 0} S(x_1,y_1,...,y_{m-1},x_m)\equiv S_m =\alpha + \beta m
\,,  \label{limite}
\end{equation}
where 
\begin{eqnarray}
\alpha=2\gamma-\delta\,, \\
\beta=\delta -\gamma \,.
\end{eqnarray}
For $m=1$ and $m=2$ we have proved this formula. Suppose it is
valid for $m\le
n $. From the constructions of fig.(\ref{unid})c and fig.(\ref{unid})d and SSA we have 
\begin{eqnarray}
S(x_1,y_1,...,y_{n-1},x_n)+S(x_n, y_n, x_{n+1})&\ge&
S(x_1,y_1,...,y_{n},x_{n+1})+ S(x_n) \,, \\
S(x_1,y_1,...,y_{n},x_{n+1})+S(x_n+y_n+x_{n+1})&\ge& \\
S(x_n,y_{n},x_{n+1})&+& S(x_1,y_1,...,y_{n-1},x_n+y_n+x_{n+1}) \,,
\end{eqnarray}
respectively.
Taking the limit of small $x_1+y_1+...+y_n+x_{n+1}$ in these inequalities and using the induction hypothesis it
follows that (\ref{limite}) is also valid for $m=n+1$. Then it holds for any
integer $m\ge 1$.

It is immediate from subadditivity and from (\ref{limite}) that
\begin{equation}
0\le\gamma\le\delta\le 2\gamma \,.
\end{equation}
These relations for the parameters simply translates into the positivity
of the independent variables $\alpha$ and $\beta$,
which do not have to satisfy any constrain.
We will also write (\ref{limite}) as
\begin{equation}
S_m=\gamma+(m-1)\beta\,,  \label{ootra}
\end{equation}
where $\gamma\ge \beta\ge 0$.

\subsection{The multidimensional case}

Consider now a translation invariant state in ${\bf R}^{d}$. Let the unit
vectors $v_1, ..., v_d$ form an orthogonal basis of ${\bf R}^d$.
Define the
family of rectangular polyhedra as all the translates of the
$R(a_{1},...,a_{d})=\{x/
x=\sum_{i=1}^d \lambda_i v_i, \lambda_i\in [0,a_i]\}$, where $
(a_{1},...,a_{d})$ are the sides lengths. We often use just a symbol $R$ for 
these polyhedra, what should not be confused with the symbol for the reals ${\bf R}$.
  The entropy can be written as a function on $d$ variables $
S(a_{1},...,a_{d})$ on these sets. The same constructions made above for the
one dimensional case apply here on each direction separately. Therefore $
S(a_{1},...,a_{d})$ is a non decreasing and concave function in each $a_{k}$.
Then the entropy is ordered by inclusion among the rectangular polyhedra.
Also, by
concavity $S((a_{1},...,a_{d})/(a_{1}\times ...\times a_{d})$ is non
increasing in each $a_{k}$ separately. Then the entropy divided the $d$
dimensional volume for
 has a limit for the rectangular polyhedra when the volume goes to infinity along any
given increasing sequence $R_{n}$. When all the sides of $R_{n}$ go to
infinity, the limit $s$ of entropy over volume must be the same number
independently of the particular sequence. This is because giving two of such
sequences of rectangular polyhedra with sides going to infinity, every element
of one sequence must be smaller than some element of the other sequence, and
in consequence none of the limits of the entropy over volume could be bigger
than the other. Thus $s$ is really the entropy per unit $d$ dimensional
volume of the system. If the volume goes to infinity but one of the sides
remains bounded the limit of entropy per unit volume could differ from $s$.
Physically, this means that the effects of the boundary may always be
relevant in this case. 

When in addition to symmetry under translations there
is rotational symmetry this result can be generalized to different sequences
of polyhedra. Any sequence  where the
volume goes to infinity at a greater rate than the bounding area (going to
infinity in the sense of Van Hove), and where the number of faces remains
bounded, has the limit $s$ of entropy over volume, which is independent of
the particular sequence \cite{Wehrl}. It is in this type of sequences that one expects the
surface terms contribution to the entropy to be negligible in the limit. We
do not use this result here. Instead we are more interested in the
properties of the entropy function for sets with finite area and vanishing
volume than in the infinite volume limit. These flat sets may not be so
interesting in statistical mechanics but they are crucial for the
relativistic geometric entropy. The boost symmetry 
 allows to amplify their width and reduce most of the general relativistic problem 
to the case of flat Euclidean sets. 

\subsection{Entropy for flat sets}

Let us consider polyhedra on ${\bf R}^{d}$ of the form $X\times
[x_{1},y_{1},...,y_{m-1},x_{m}]$, where $X$ is a $d-1$ dimensional polyedron
 and $[x_{1},y_{1},...,y_{m-1},x_{m}]$ is a set with $m$
connected components in ${\bf R}^1$.
 Here $x_1,...,x_m$ are the single component lengths and $y_1,...,y_{m-1}$ are
 the separation distances between adjacent components. Call the corresponding entropies
$S(X,x_{1},y_{1},...,x_{m})$. For a fixed $X$ these polyhedra form a
translation invariant one dimensional system, and we have from
equation (\ref{ootra}) that
\begin{equation}
\lim_{x_{1}+y_{1}+...+x_{m}\rightarrow 0}S(X,x_{1},y_{1},...,x_{m})=\gamma
(X)+ (m-1)\,\beta (X)\,,
\end{equation}
where $\gamma (X)$ and $\beta (X)$ are positive and Euclidean invariant
functions on polyhedra in ${\bf R}^{d-1}$, and $\gamma (X)\ge \beta (X)$. It
follows from SSA and SSB in $d$ dimensions that $\gamma (X)$ and $
\beta (X)$ also satisfy SSA and SSB. Note that we use the same symbols 
$\beta$ and $\gamma$ that represent constants in the one dimensional case 
for denoting functions in the multidimensional case.

\begin{figure}
\centering
\leavevmode
\epsfysize=4cm
\epsfbox{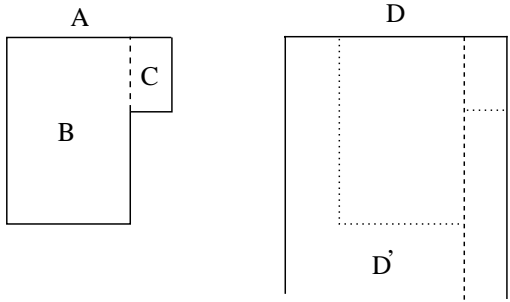}
\bigskip
\caption{Eating from inside a strip $D$ to a smaller strip $D^\prime$
 (the right side of $D^\prime$ is the dashed line), by means of
 $A-B$. One has to position $A$ inside $D$ (shown with the dotted line) and take out from
 $D$ the sector corresponding to $A-B=C$. Then translating down $A$ inside
 $D$ and repeating the operation it is possible to eat all $D-D^\prime$.}
\label{ea}
\end{figure}

In addition, these Euclidean entropy functions are also ordered by inclusion. To show it we first
introduce a
method that we can call the 'eating from inside' procedure. This was
developed
 in \cite{rob} as a tool to prove certain monotonicity relations.

Suppose we have two sets $A$ and $B$ with $B\subset A$, and we want to
obtain information about the difference of their entropies. Call $C=A-B$. At
our convenience we choose a set $D$ with $A\subset D$. Then from strong
subadditivity it follows that $S(A)+S(D-C)\ge S(D)+S(B)$. This can be
written as $S(A)-S(B)\ge S(D)-S(D-C)$. Now, if a symmetric copy $A_1$
of $A$ can
be situated inside $D-C$ we can repeat the procedure with $D-C$ in the place of $C$, leading to $
S(A)-S(B)\ge S(D-C)-S(D-C-C_{1})$, where $C_{1}$ is the symmetric copy of $
C $ corresponding to the same transformation that carries $A$ into $A_1$.
Summing up this and the previous inequality we have that $
2(S(A)-S(B))\ge S(D)-S(D-C-C_{1})$ . The procedure can be continued if
 at the $n^{th}$ step
 we can place a copy of $A$ inside  what
is left from $D$ after subtracting all the previous symmetric the copies
of $C$ (see fig.(\ref{ea})). In that case
we have
\begin{equation}
S(A)-S(B)\ge \frac{1}{n}\left( S(D)-S(D^{\prime })\right) \,,\label{eating}
\end{equation}
where $D^{\prime }=D-C-C_{1}-...-C_{n-1}$. In some situations the right-hand side
in (\ref{eating}) is known, specially if the limit $n\rightarrow \infty $
can be taken. From this we can obtain information about the difference of
entropies of two sets $A$ and $B$ ordered by inclusion. We say
that to obtain (\ref{eating}) we eat from inside $D$ to $D^\prime$ by means
of $A-B$.

Take now any $d-1$ dimensional polyhedra $X$ and $Y$ with $Y\subseteq X$, and
let the one dimensional set $Z=[x_{1},y_{1},...,y_{m-1},x_{m}]=[a,a,...,a,a]$
be formed by $m$ intervals of length $a$, with separation distance $a$
 between adjacent intervals.
 Consider the sets $A=X\times Z$, $B=Y\times Z$, $
C=(X-Y)\times Z$, and $D=X\times [2maN]$. Here $[l]$ means a single interval
of length $l$, and $N$ is a big integer. Then it is easy to see that
translating $A$, $B$, and $C$ on the direction  perpendicular to
$X$ and $Y$ we can eat from inside $2N$ copies
of $C$ from $D$, leading to the remaining set $D^{\prime }=Y\times [2maN]$.
Application of (\ref{eating}) gives 
\begin{equation}
S(A)-S(B)\ge \frac{ma}{(2maN)}\left( S(D)-S(D^{\prime })\right) \,.
\end{equation}
The results for one dimensional translation invariant systems imply
that $S(U\times [l])/l$ has a limit $s_{U}$ for $l$ going to infinity. This
implies taking $N\rightarrow \infty $ that 
\begin{equation}
S(A)-S(B)\ge ma\,(s_{X}-s_{Y})\,.
\end{equation}
Now, taking the limit $a\rightarrow 0$ we have from this inequality that $\gamma (X)+(m-1)\beta
(X)$ is ordered by inclusion for any $m\ge 1$. Consequently $\gamma (X)$ and 
$\beta (X)$ are ordered by inclusion.

\begin{figure}
\centering
\leavevmode
\epsfysize=4cm
\epsfbox{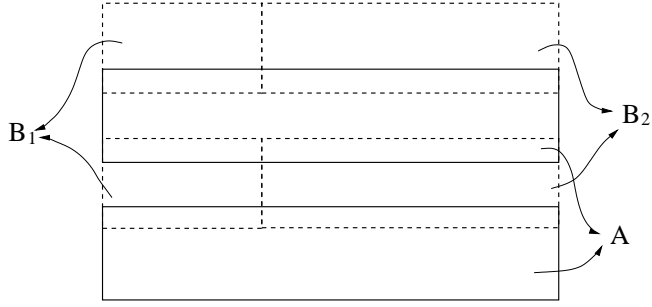}
\bigskip
\caption{The geometric construction used to show that the function $\beta$ 
is proportional to the volume.
 The solid line represents a transversal view of the two component set $A$. Drawn with
dashed lines are the two component set $B_1$ on the left and the two component set $B_2$
on the right. All the connected components of these sets have the same height $a$.}
\label{betavo}
\end{figure}

Moreover, the functional form of $\beta (X) $ can be calculated explicitly.
Let us take
$A=X\times Z$ with $Z=[x_{1},y_{1},x_{2}]=[a,a/2,a]$ and $X$ a $d-1$
dimensional polyhedron. Also take $B_{1}=Y_{1}\times Z$ and $B_{2}=Y_{2}\times
Z$ with $Y_{1}$ and $Y_{2}$ two $d-1$ dimensional polyhedra such that $
X=Y_{1}\cup Y_{2}$ and $Y_1\cap Y_2 =\emptyset$. Suitably placing $B_{1}$ and $B_{2}$ to have empty
 (or measure zero) 
intersection and to cover the gaps on $A$ (see fig.(\ref{betavo})), we
can arrange that the union $A\cup B_{1}\cup B_{2}=X\times [(13/4)a]$ and the
intersections $A\cap B_{1}=Y_{1}\times Z^{\prime }$, $A\cap
B_{2}=Y_{2}\times Z^{\prime }$, where $Z^{\prime
}=[x_{1},y_{1},x_{2},y_{2},x_{3}]=[a/4,a/2,a/4,a/2,a/4]$. Then, from strong
subadditivity for the pairs $(A,B_1)$ and $(A\cup B_1,B_2)$, and taking the 
limit $a\rightarrow 0$ we get 
\begin{equation}
\beta (X)\ge \beta (Y_{1})+\beta (Y_{2})\,.
\end{equation}
This inequality is opposite to the subadditive inequality for $\beta $ since 
$X=Y_{1}\cup Y_{2}$. Thus the equation holds, 
\begin{equation}
\beta (X)=\beta (Y_{1})+\beta (Y_{2})\,.  \label{tere}
\end{equation}
With $N^{d-1}$  copies of a rectangular polyhedra $R$ with
sides $(l_{1},...,l_{d-1})$ we can exactly cover a rectangular polyhedron $U$
with sides $(N\,l_{1},...,N\,l_{d-1})$. Using (\ref{tere}) it is
\begin{equation}
\beta (R)=\text{vol}(R)\,\frac{S(U)}{\text{vol}(U)}\,.
\end{equation}
The fraction on the right hand side converges to a limit $2s$ for large $N$.
 The factor $2$ is chosen for future convenience. 
Thus, we have that the function $\beta $ is proportional to the volume for
any rectangular polyhedron. To show that $\beta (X)$ is proportional to the
volume for any polyhedron $X$ divide the space ${\bf R}^{d-1}$ with a cubic
grid formed by copies of the cubic element
 $A$. Call $\Gamma ^{-}(A,X)$ the union of elements of the grid that
are included in $X$, and $\Gamma ^{+}(A,X)$ the union of elements of the
grid having non empty intersection with $X$. Using (\ref{tere}) and that $
\beta $ is proportional to the volume on cubes it follows that the entropies
 $S(\Gamma ^{\pm}(A,X))=2s\,\text{vol}
 (\Gamma ^{\pm}(A,X))$ are proportional to the volumes. Then, using
 the order by inclusion of $\beta$, the fact that $\text{vol}(\Gamma
^{-}(A,X))\leq \text{vol}(X)\leq \text{vol}(\Gamma ^{+}(A,X))$, and that these volumes
coincide in the limit of small $A$, we obtain
\begin{equation}
\beta (X)=2s\, \text{vol}(X)
\end{equation}
 for every $X$. Note that the only symmetry we used was translational
invariance.

The function $\alpha =\gamma -\beta $ is positive and strong subadditive
because $\beta(X)=2s\,\text{vol}(X) $  satisfies the strong
additive equality, $\beta(A)+\beta(B)=\beta(A\cap B)+\beta(A\cup B)$.

Summarizing, we have shown the following theorem on the entropy limit
for flat sets

\vspace{.3cm}

{\bf Theorem 1:} For an Euclidean invariant $d$ dimensional system
 the limit of the entropy for sets of the type
 $X\times [x_{1},y_{1},...,x_{m}]$
as all the $x_{1},y_{1},...,x_{m}$ go to zero exists and is given by
\begin{equation} 
\lim_{x_{1}+y_{1}+...+x_{m}\rightarrow 0}S(X,x_{1},y_{1},...,x_{m})=\gamma
(X)+(m-1)\,2s\,\text{vol}(X)\,\,,
\end{equation}
where $\gamma (X)$ is an ordered by
inclusion $d-1$ dimensional Euclidean entropy function,
and $\gamma (X)\ge 2s\,$vol$(X)$.

\vspace{.3cm}

One could wonder about the entropy limits for lower dimensional objects.
Define $\hat{\gamma}$ and $\hat{\beta}$ as the
operators that transform Euclidean $d$ dimensional entropy
functions into Euclidean $d-1$ dimensional entropy functions
 and whose action is defined as  $\hat{\gamma}_S=\hat{\gamma}(S)=\gamma$,
 $\hat{\beta}_S=\hat{\beta}(S)=\beta$, with $\gamma$ and $\beta$ as above.
  The monotonicity of $\gamma$ and $\beta$ imply 
that $\hat{\beta} \circ \hat{\gamma} =0$,
and $\hat{\beta} \circ \hat{\beta} =0$, and the explicit
form of $\beta $ leads to $\hat{\gamma} \circ \hat{\beta} =0$.
 However, the composition $
\hat{\gamma} \circ \hat{\gamma} $ can be different from
zero when applied to a
general entropy function $S$. We call $\hat{\gamma}^n_S$ for $n=0,...,d$ the composition of
 $S$ with $n$ times the operator $\hat{\gamma}$, and define $\hat{\gamma}^0_S=S$.

\section{Geometric entropy in Minkowski space}

\subsection{The $1+1$ dimensional case}

Before embarking in the calculation of the functional form
of the entropy  in the general $d+1$ dimensional Minkowski space
 we focus on the simpler $1+1$ case. This allows to see without the
  complications of higher dimensions the important role played by
  the boost symmetries. The result for the entropy and the method
  to obtain it are similar
 to the ones for the $1$ dimensional
Euclidean small sets.  

\begin{figure}
\centering
\leavevmode
\epsfysize=5cm
\epsfbox{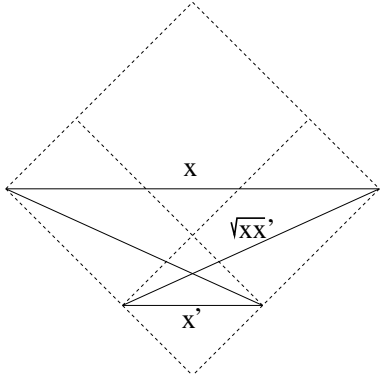}
\bigskip
\caption{Two commuting sets with straight Cauchy surface size $\sqrt{x x^\prime}$ 
in $1+1$ dimensions intersect in another diamond-like set of size $x^\prime$, and their union is, 
after causal completion, a diamond set of base size $x$.}
\label{umasu}
\end{figure}

The connected causally closed sets are determined in $1+1$ dimensions by a single
number $x$, the size of the unique straight Cauchy surface (the size of the diamond base, see 
fig.(\ref{umasu})). 
The corresponding entropy $S(x)$
is a positive, non decreasing and concave function because of translation
symmetry. As above, call $\gamma$ the constant $\lim_{x\rightarrow 0}S(x)$.

The additional boost symmetry greatly constrain the function $S(x)$. To see
this, consider the construction of fig.(\ref{umasu}). Given $x^{\prime} \leq x$ we
place two commuting boosted
one component sets of size $\sqrt{x\,x^{\prime}}$ in such a way
their
intersection is a one component set of size $x^{\prime}$ and their union is another one component set 
of size $x$. The application of SSA gives
\begin{equation}
S(x)+S(x^{\prime})\leq 2\, S(\sqrt{x\,x^{\prime}})\,.
\end{equation}
Making $x^{\prime}$ as small as we want in this inequality
we prove that $S(x)\leq \gamma$. As $S(x)$ is
non decreasing this means that $S(x)\equiv \gamma$ is constant. Due to the
boost symmetry we can form a Cauchy surface for a given diamond $D$ using
two boosted diamonds of size as small as we want disposed along the null
surfaces of $D$. This, and the subadditive law, is what impedes an entropy 
 increasing like the volume.

The entropy for sets consisting on several components spatially separated can
in principle depend on the lengths of the individual diamond bases, their
separation distances, and relative angles.

\begin{figure}[t]
\centering
\leavevmode
\epsfxsize=12cm
\epsfbox{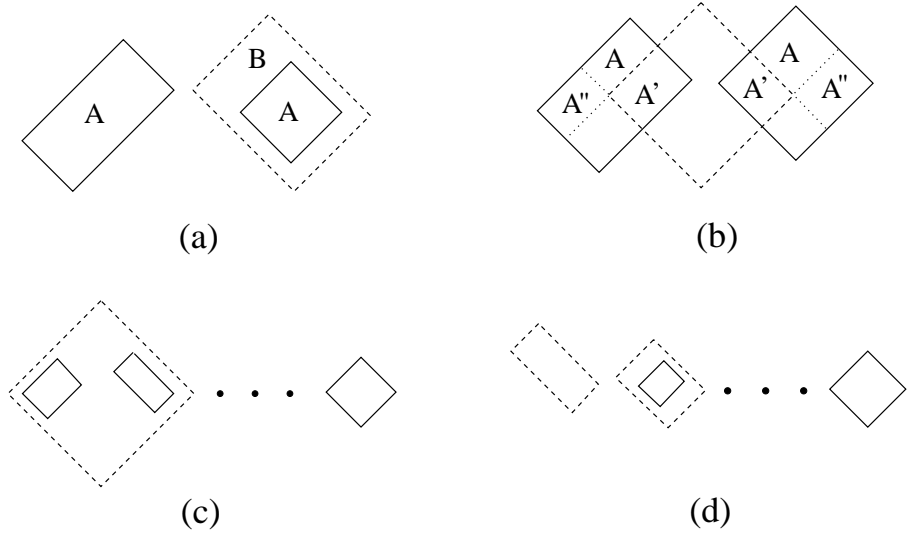}
\bigskip
\caption{Two dimensional constructions used to prove Theorem 2.
 In (a), (c) and (d) it is considered the SSA between the solid line and the dashed line set.
 In (b) both the SSA and the SSB between the solid line set and the central dashed diamond
are used.}
\label{umasub}
\end{figure}

Let us start with the case of two component sets. In
the construction of fig.(\ref{umasub})a the union of a two component set $A$
with a single diamond $B$ is a set $A^{\prime}$, having two components, one
of which is bigger than the corresponding component in $A$.
 Application of SSA gives
\begin{equation}
S(A)+\gamma\ge S(A^{\prime})+\gamma\,.
\end{equation}
Thus, the entropy decreases with increasing size of the individual connected components.
Now, the application of SSA and SSB to the construction of
fig.(\ref{umasub})b leads to the inequalities 
\begin{eqnarray}
S(A)+\gamma & \ge & S(A^{\prime})+\gamma\,, \\
S(A)+\gamma & \ge  & S(A^{\prime\prime})+\gamma\,,
\end{eqnarray}
where $A^{\prime}$ and $A^{\prime\prime}$ are both included in $A$ but $
A^{\prime}$ shares with $A$ the internal spatial corners and $
A^{\prime\prime}$ shares with $A$ the external spatial corners. Combining
these relations we can prove that the entropy of any two component
set $X$
is greater than the entropy of any other two component set $X^{\prime}$ where
each individual component of $X^{\prime}$ is included in a corresponding
component of $X$. Thus, the entropy of the two component sets is increasing and
decreasing with inclusion, and then it must be a constant. Call $
\delta $ this constant.

For a greater number of connected components the entropy can be expressed in
terms of $\gamma$ and $\delta$ and we show the following

\vspace{.3cm}
{\bf Theorem 2:} The most general form for a relativistic entropy function in $1+1$
dimensional Minkowski space is given by
\begin{equation}
S(Z)\equiv S_m=\gamma+(m-1)\, \beta\,,  \label{unidi}
\end{equation}
where $m$ is the number of connected components of $Z$ and $\gamma\ge \beta\ge 0$.

\vspace{.3cm}
In terms of $\delta$ it is $\beta=\delta-\gamma$. 
The proof
is very similar to the corresponding one for the small one dimensional sets
in Section III. To show (\ref{unidi}) by induction on the number of
components note that for $m=1$ and $m=2$ this equation holds. Supposing it
is valid for $m\le
n $ we have from the constructions of fig.(\ref{umasub})c and fig.(\ref{umasub})d 
\begin{eqnarray}
S_{n+1}(X)+\gamma \ge S_n +\delta\,, \\
S_n+\delta\ge S_{n+1}(X)+\gamma\,.
\end{eqnarray}
This proves the formula (\ref{unidi}). It follows from positivity and
subadditivity that the parameters have to satisfy the relations
\begin{eqnarray}
0\leq \gamma \leq \delta \leq 2\, \gamma \,, \\
0\le \beta\le \gamma\,. \label{relatio}
\end{eqnarray}
The meaning of these constraints is clarified
if we write $S_m$
as a sum 
\begin{equation}
S_m=\alpha+ \beta\, m\,,  \label{euler}
\end{equation}
of two independent solutions with independent and arbitrary positive
coefficients. Here
$\alpha=\gamma-\beta$. One of the
solutions is the constant term and the other is topological, being
proportional to the number of connected components $m$. It is easy to
check that
(\ref{euler}) does indeed satisfy SSA and SSB for two arbitrary commuting
sets. Thus, this formula, or eq. (\ref{unidi}), represents the most general solution for the entropy
 in $1+1$ dimensional Minkowski space.

\subsection{Multidimensional case $d>1$}

In this Section we show  the theorem

\vspace{.3cm}
{\bf Theorem 3:} The most general form for a relativistic  entropy function 
on the relativistic polyhedra on $d+1$ dimensional  Minkowski space, with $d>1$, is given by 
\begin{equation}
S(X)=\alpha _{0}+s\,\,\text{area}(X)\,,
\end{equation}
where $\alpha_0$ and $s$ are non negative constants and $\text{area}(X)$ is the area of the
 spatial corner of $X$.

\vspace{.3cm}
 We divide the proof in several parts.

\subsubsection{Relativistic Lemmas}

To apply SSA or SSB
between two sets we have to ensure that they commute, what
complicates the proofs for $d>1$.
In the following we incorporate into the Euclidean
 function given by restricting the relativistic entropy function to a single 
spatial hyperplane some information coming from the boost symmetry. With these 
 additional properties at hand we will be able to work in Euclidean space where 
all sets commute.

 An example where commutativity fails is given by adding one perpendicular 
spatial dimension to the sets in figure (\ref{umasu}). The two boosted intervals of length 
$\sqrt{x\,x^\prime}$ there can be seen as the transversal section of two boosted 
rectangles
 in $2+1$ dimensional Minkowski space. However,  these 
 rectangles do not commute when they are
 in such a position since their  
boundaries are not spatial to each other. 
 
There is a limit where commutativity can be applied in an effective form. Consider
 the product $R\times Z$ of a big $d-1$ dimensional rectangular polyhedron $R$ and a 
relativistic polyhedron $Z$ in $1+1$. We mean here by the product of a polyhedron $R$
 with a
relativistic polyhedron $Z$ the relativistic polyhedron generated by the 
Cartesian product of $R$ with a Cauchy surface for $Z$, since the Cartesian product of $R$ 
with the relativistic polyhedron corresponding to $Z$ is not a relativistic polyhedron 
on $d+1$.
Given two commuting $1+1$ dimensional sets $Z$ and $Z^\prime$ the sets 
$R\times Z$ and $R\times Z^\prime$ 
may not commute. However, 
taking two fixed commuting $Z$ and $Z^\prime$, 
 we can construct a relativistic 
 polyhedron $U$ commuting with $R\times Z$ by deforming $R\times Z^\prime$ near the borders of $R$. 
More precisely, we take $U$ as the domain of dependence of  
 the union of $(R\times Z^\prime)\cap (R\times Z)$ and $(R\times Z^\prime)\cap (R\times Z)^\perp$.  
Here $A^\perp$ is the set formed by all the points spatially separated from $A$.  
When the sides of $R$ are big, $U$ differs from $R\times Z^\prime$ by a set 
located along the boundary of $R$, and whose transversal size remains
 fixed if the sizes of $R$ are increased. Thus, the difference in entropy
between $R\times Z^\prime$ and $U$ can be bounded by a 
number proportional to the border area of $R$ by using intersections and unions of 
rectangular polyhedra with bounded sides. 
If we consider the limit of the entropy $S(R\times Z)$ divided by
 the volume $\text{vol}(R)$ 
for the sets of the type $R\times Z$ with $R$ going to infinity the terms growing like the 
  border area of $R$ can be neglected. Therefore, in this limit we 
can use commutativity between the sets $R\times Z$ and $R\times Z^\prime$ with commuting $Z$ and $Z^\prime$, 
and use the two dimensional results. We have the 

\vspace{.3cm}
{\bf Lemma 1:} Let $R$ be a $d-1$ dimensional rectangular polyhedron and $Z$ a $1+1$
relativistic polyhedron then 
\begin{equation}
\lim_{R\rightarrow \infty} \frac{S(R\times Z)}{\text{vol}(R)}=\gamma_\infty +2s\, (m-1) \,,
\end{equation}
where $m$ is the number of connected components in $Z$, $s$ is defined by 
Theorem 1, and $\gamma_\infty$, with $\gamma_\infty \ge 2s$, is the 
limit $\gamma(R)/\text{vol}(R)$ for $R$ going to infinity.

\vspace{.3cm}

The explicit form of the right hand side follows from Theorem 1, since the two dimensional
 solution tell us that this term must be a constant plus a constant times
 the number of components of $Z$. Then we can calibrate the parameters from the special 
case of $Z$ lying in a single spatial line, $Z=[x,y,...,y,x]$, for small $x$, $y$.

In particular the Lemma applies to the sets  
$Z=[x_1,y_1,...,x_m]$ that lie in a single spatial plane, with arbitrary  $x_1, y_1,..., x_m$. 
Thus, it removes the constrain of small $x_1+y_1+...+x_m$ in Theorem 1, in the limit of $R$ going to infinity. 
Of course, this includes information about boost symmetry and it is not the case 
for a general Euclidean function.

\begin{figure}
\centering
\leavevmode
\epsfysize=4cm
\epsfbox{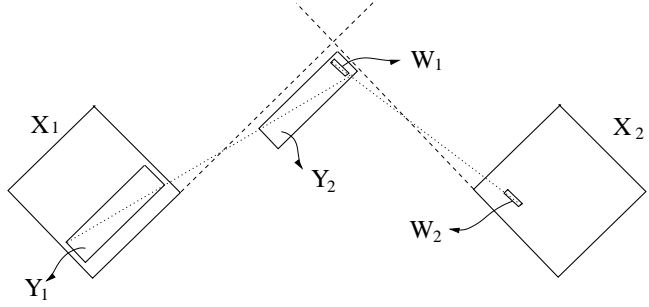}
\bigskip
\caption{Two dimensional cut of the geometrical construction used to prove inequality (\ref{cinctre}).
 The lines at $45^\circ$ are null surfaces. The dotted lines represent the spatial hyperplanes
containing the defining Cauchy surfaces for $Y=Y_1\cup Y_2$ and $W=W_1\cup W_2$.}
\label{peque}
\end{figure}

With the notation of Section IV consider now the entropies $S(R,x_1,y,x_2)$
 for the sets $X=R\times [x_1,y,x_2]$ contained in a single spatial hyperplane, 
and where $R$ is a fixed $d-1$ dimensional rectangular polyhedron. 
Call $X_1=R\times [x_1]$ and $X_2=R\times [x_2]$. Theorem 1 showed the form
of this function for the limit in which all $x_1,y,x_2$ go to zero.
We can do better in the relativistic case, and find the limit for $x_1$ and $x_2$ 
going to zero and any $y$.
The first inequality we obtain comes eating from inside  
$U\times [x_1,y,x_2]$, where $U$ is a big rectangular polyhedron, 
to $U\times [x_1]$ by means of $X-X_1$. Using the  Lemma 1 and Theorem 1 
 this leads to
\begin{equation}
\lim_{x_1+x_2\rightarrow 0} S(R\times [x_1,y,x_2])\ge \gamma(R)+ 2s\, \text{vol}(R)\,.\label{inee}
\end{equation}
Then consider the following construction shown in the figure (\ref{peque})
 through a two dimensional projection.
The set $Y=Y_1\cup Y_2$ is of the form $R^\prime\times [x_1^\prime,y^\prime,x_2^\prime]$,
but it appears highly boosted with respect to $X$ in the direction perpendicular to $R$.
The same happens for $W=W_1\cup W_2=R^{\prime\prime}\times[x_1^{\prime\prime},y^{\prime\prime}
,x_2^{\prime\prime}]$. The rectangular polyhedron $R^{\prime\prime} \subseteq R^{\prime}\subseteq R$,
the transversal
sizes $(x_1^\prime,y^\prime, x_2^\prime, x_1^{\prime\prime}, y^{\prime\prime},x_2^{\prime\prime})$, 
 and the boost factors are chosen in such a way that $Y_1\subseteq X_1$, $W_1\subseteq Y_2$,
$W_2\subseteq X_2$, and $Y_2$ is spatial to $X_1$ and $X_2$. 
We are taking then the limit of small 
$(x_1,x_2,x_1^\prime,y^\prime,x_2^\prime,x_1^{\prime\prime},y^{\prime\prime},x_2^{\prime\prime})$
 but keeping $y$ fixed. In this limit the sides of $R^\prime$ and $R^{\prime\prime}$ can be taken to
converge to those of $R$.  
Applying successively SSA to the pairs of sets $(X_1,Y)$, $(X_1\cup Y_2,W)$, and 
$(X_1\cup Y_2 \cup W_2,X_2)$ we have
\begin{equation}
S(X\cup Y_2)\le 2\gamma(R)-\gamma(R^{\prime\prime})+2s\, \left( \text{vol}(R^\prime) +
\text{vol}(R^{\prime\prime})\right)\,+\epsilon\,.\label{cinctre}
\end{equation}  
for $\epsilon$ as small as we want. 
Now, we eat from inside a three component set formed by the product of a big rectangular $d-1$ dimensional
set $U$ and the relativistic $1+1$ set given by the transversal projection of $X\cup Y_2$ 
(the set shown in figure (\ref{peque})) to $U\times [x_1,y,x_2]$ by means of $(X\cup Y_2)-X$. Taking 
into account Lemma 1 this leads to
\begin{equation}
S(X\cup Y_2)-S(X)\ge 2s\, \text{vol} (R^\prime)\,.
\end{equation} 
Combining the last two equations, taking a limit 
in which $R^{\prime\prime}$, $ R^{\prime}$, and $ R$ all have the same volume and shape, and
using the continuity of $\gamma$ as a function of the sides of the rectangular polyhedron, we
obtain the opposite inequality to (\ref{inee}). Then, we have
\begin{equation}
\lim_{x_1+x_2\rightarrow 0} S(R\times [x_1,y,x_2])= \gamma(R)+2s\, \text{vol}(R)\,.\label{qq}
\end{equation}

This equation allows us to prove the following 
 
\begin{figure}
\centering
\leavevmode
\epsfysize=4cm
\epsfbox{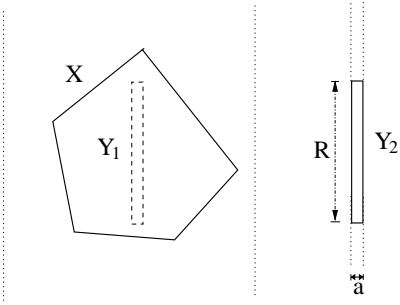}
\bigskip
\caption{The entropy of $X\cup Y_2$ is equal to $S(X)+2s\, \text{vol}(R)$ in the limit
 of small $a$. The dotted lines represent a two component strip-like set that can be eaten
 from inside down to a one component strip by means of $((X\cup Y_2)-X)$.}
\label{linea}
\end{figure}

 \vspace{.3cm}
{\bf Lemma 2:} Let $X$ be a $d$ dimensional polyhedron, $R$ a $d-1$ dimensional
 rectangular polyhedron,  $Z=[a,y,a]$, $Y=R\times Z$, and $Y_1$ and $Y_2$
 the connected components of $Y$, symmetric copies of $R\times [a]$. 
Suppose that $Y\cap X=Y_1$ and that there is an
 open set $O$ including $Y_2$ such that any translation of $O$ in a plane parallel to $R$
 has no intersection with $X$ (see fig.(\ref{linea})). Then
\begin{equation}
\lim_{a\rightarrow 0} S(X\cup Y_2) =S(X) + 2s\, \text{vol}(R)\,.
\end{equation}

\bigskip
The proof uses the previous result (\ref{qq}) and Lemma 1. 
Applying SSA to the pair $(X,Y)$ we have
\begin{equation}
 \lim_{a\rightarrow 0}S(X\cup Y_2) \le S(X) + 2s\,\text{vol}(R)\,.
\end{equation}  
The opposite inequality follows from eating from inside a two component large 
rectangular strip-like set,
 one component containing $X$, the other having transversal size equal to $a$ and containing $Y_2$, as
show in fig.(\ref{linea}).

\subsubsection{Subtraction of the constant term}

Here we show that the limit $\alpha_0$ of the entropy for rectangular
polyhedra with vanishing small sides can be identified with the
constant solution component of the general solution. In other words
 we show the following

\vspace{.3cm}

{\bf Proposition 1: } Let $S$ be a relativistic entropy function. Then 
$S=\alpha_0+S^1$, where $\alpha_0$ is the limit value of the entropy for 
rectangular polyhedra of vanishing small sides, and $S^1$ is a relativistic 
entropy function with zero entropy limit for the rectangular polyhedra 
with vanishing small sides. 

\vspace{.3cm}
We know that the general solution must have a constant term. However, the 
identification of it with $\alpha_0$ and 
the subtraction of $\alpha_0$ are not automatic, since we have to show that $
S^{1}$ is still positive and satisfies SSA and SSB. An example where
 this is not the case is given by the two dimensional solution (\ref{unidi}).
 If one subtracts the 
entropy limit for small intervals ($\gamma$ in the two dimensional case), 
the result is a positive
 but not strongly subadditive function. This is because subtracting 
$\gamma$ is different from subtracting the constant solution. There is 
a non constant solution, the one proportional to $\beta$, that is constant over 
the one component sets and adds
 to $\gamma$. However, as it will be clear from what follows, 
 in more dimensions there is no such solution. In particular, there are no positive 
topological solutions to the strong subadditive inequalities.  
Curiously, the  Euler number $E(X)$, which is the second solution appearing in 
the two dimensional formula (\ref{euler}), satisfies the strong additive equality,
 $E(A)+E(B)=E(A\cup B)+E(A\cap B)$ in any dimension. However, for $d>1$ it attains 
negative values.

\begin{figure}[t]
\centering
\leavevmode
\epsfxsize=12cm
\epsfbox{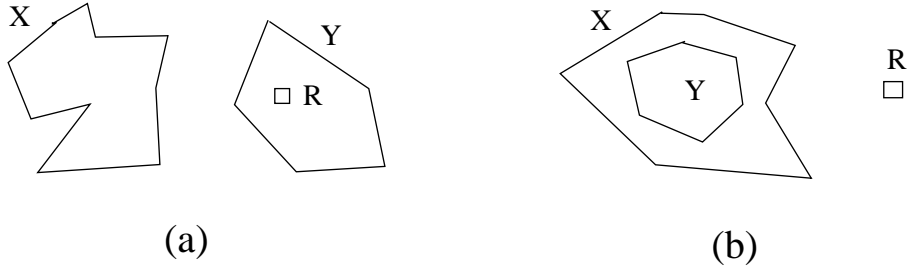}
\bigskip
\caption{(a)- The entropies of two non intersecting sets $X$ and $Y$ satisfy the relation (\ref{eee}).
 This is shown by constructing $X^\prime=X\cup R$ and using SSA on the pair $(X^\prime,Y)$.
 (b)- For $Y\subseteq X$ we can show that $S(X)+S(Y)\ge S(X-Y)+\alpha_0/2$ by using SSB on the 
pair $(X^\prime,Y)$, where $X^\prime=X\cup R$ and $R$ is a small rectangular polyhedron  
 spatially separated from $X$.}
\label{sus}
\end{figure}

Now consider any relativistic polyhedron $X$. It contains a sufficiently small
 rectangular polyhedron $R^\prime$ lying in a spatial hyperplane ${\cal P}$.
We can generate a copy $R$ of $R^\prime$ by translating $R^\prime$ on ${\cal P}$ in
a direction perpendicular to a phase $F$ of $R^\prime$, and such that $R\cap X=\emptyset$.
 For $R$ sufficiently far from $X$ there is a time-like hyperplane parallel 
to $F$ in
between $X$ and $R$ that divides the Minkowski space in two parts, one
 of which contains $X$ and the other $R$. Let us call $X^\prime=X\cup R$ (see fig.(\ref{sus})a).
Now, from Lemma 2 we have that for small enough $R$
\begin{equation}
S(X^\prime)-\epsilon\le S(X)\le S(X^\prime)+\epsilon\,. \label{dosde}
\end{equation} 
Then, by means of a $\pi$ angle rotation and a small translation on ${\cal P}$ we can make 
a copy of $X^\prime$ whose intersection with $X^\prime$ is a small rectangular polyhedron inside $R$.
It follows from strong subadditivity that $S(X^{\prime})\ge (1/2)\, \alpha_0$. 
Then taking the limit of small $R$ we have
\begin{equation}
S(X)\ge \frac{1}{2} \alpha_0
\end{equation}
for any $X$.

 Thus, we can safely subtract $(1/2) \alpha_0$
from the entropy without violating positivity. Strong subadditivity for the
subtracted function will be automatic for the case where the two commuting
sets $X$ and $Y$ have non empty intersection. When the intersection $X\cap Y=\emptyset $ 
we have to prove 
\begin{equation}
S(X)+S(Y)\ge S(X\cup Y)+\frac {\alpha_0}{2}\,, \label{eee}
\end{equation}
to show that the subtracted function is still subadditive. To see this, 
consider a set $X^\prime=X\cup R$ as above, but now we choose 
$R$ included in $Y$ (see fig(\ref{sus})a). Then 
\begin{equation}
S(X^\prime)+S(Y)\ge S(X\cup Y)+S(R)\,.
\end{equation}
From (\ref{dosde}) and taking limits the relation (\ref{eee}) follows.

 This shows that subtracting the constant solution $\alpha_0/2$ leads to 
a positive strongly subadditive solution. It also satisfies SSB
 as follows from similar arguments based on the construction of fig(\ref{sus})b 
for the case in which the inequality is applied to the sets $X$ and $Y$ with
$Y\subseteq X$ (otherwise there are two non empty sets on each side of the
SSB inequality, which holds trivially for the subtracted function).

Once we have subtracted $\alpha_0/2$ we can repeat the procedure 
with the subtracted function $S-\alpha_0/2$ to subtract $\alpha_0/4$, 
and then $\alpha_0/8$, etc..
 It follows that
 we can subtract a constant solution $x\alpha_0$ for any $0<x<1$ and still 
get a  solution. Thus $S-\alpha_0$ is also positive and satisfies SSA and SSB. 

\subsubsection{ $\gamma(X)$ is proportional to the volume}
Taking into account the Proposition 1 we work with the subtracted function and 
 assume $\alpha_0=0$. We show that 
$\gamma (X)=\beta (X)=2s\, \text{vol}(X)$, or, what is the same, $\alpha(X)=0$.  
To do so we need to make some geometrical constructions and use that
 the entropy for sets that are flat in more than one perpendicular direction can be 
safely neglected. In the notation of Section III this writes $\hat{\gamma}^n(S)=0$ 
for $n=2,...,d$. It is possible to write this last equation and the result we are looking for, 
$\gamma (X)=\beta (X)$ in a compact manner, that is, 
$\hat{\alpha}^n_S=(\hat{\gamma}-\hat{\beta})\circ\hat{\gamma}^{n-1}_S=0$, for 
$n=1,...,d$. Here we used that $\hat{\beta}\circ\hat{\gamma}^{n-1}_S=0$ for 
$n\ge 2$. Recall that $\hat{\gamma}^0_S=S$.
 Each $\hat{\alpha}^n_S$ is a function on the polyhedra on ${\bf R}^{d-n}$,
 with $\hat{\alpha}^1_S=\alpha$ and $\hat{\alpha}^d_S=\alpha_0=0$.

Then, we show by induction that all the functions  
$\hat{\alpha}^n_S$ for $n=1,...,d$ are zero in the relativistic case
 when $\alpha_0=0$. For $n=d$ this is true. 
We show that if $\hat{\alpha}^{n+1}_S=0$ then $\hat{\alpha}^n_S=0$.

\begin{figure}[t]
\centering
\leavevmode
\epsfxsize=9cm
\epsfbox{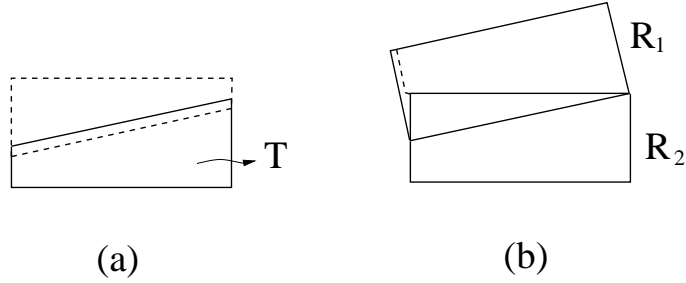}
\bigskip
\caption{Geometrical constructions used in the text to show that the limit of the entropy
 for a flat wedge-like $d-n+1$ dimensional set $T$ coincides with $\hat{\gamma}^{n}_S(R)$,
 where $R$ is the rectangular polyhedron forming  the big phase of 
$T$. The picture shows a transversal view, 
where $R$ is reduced to an interval. (a)- Two copies of $T$ are placed such that their intersection
 and union are rectangular flat polyhedra with big phase $R$ (neglecting sets small in more than one
 perpendicular direction). (b)- Two rectangular flat polyhedra $R_1$ and $R_2$ 
are placed such that both 
$R_1-R_2$ and $R_2-R_1$ are equal to $T$ modulo sets that are small in more than one direction.}
\label{tria}
\end{figure}

Consider first a polyhedra $T$ in ${\bf R}^{d-n+1}$ with a wedge-like form,
 $T=\{(x_1,...,x_{d-n+1})/ 0\le x_1\le \mu x_2 +\nu , 
0\le x_2\le a_2,..., 0\le x_{d-n+1}\le a_{d-n+1} \}$.
 Call $R$ the $d-n$ dimensional rectangular polyhedron with sides
$(a_2,...,a_{d-n+1})$.  
We are interested in the limit of $\hat{\gamma}^{n-1}_S(T)$ as $\mu$ and $\nu$ go to 
zero. Apply SSA and SSB to the constructions
of fig(\ref{tria})a and fig(\ref{tria})b respectively, and using the induction hypothesis neglect 
sets that are small in more than one direction.
We
obtain two inequalities that are opposite in the limit of zero
$\mu$, $\nu$. They lead to the equation
\begin{equation}
\lim_{\mu,\nu\rightarrow 0} \hat{\gamma}^{n-1}_S(T)=\hat{\gamma}^{n}_S(R)\,. \label{t}
\end{equation}
Thus this limit is the same as the one for flat rectangular polyhedra. In particular 
this holds for the function $S$ of flat wedge sets, what corresponds to $n=1$ in the last equation. 

\begin{figure}
\centering
\leavevmode
\epsfxsize=9cm
\epsfbox{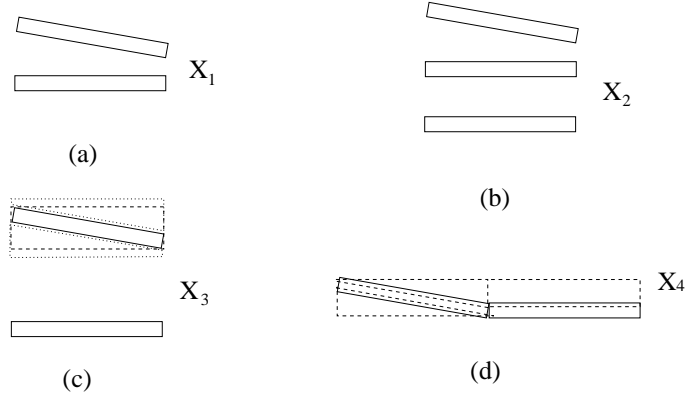}
\bigskip
\caption{The entropies of $X_1$ and $X_3$ are equal in the limit of small transversal width of their 
connected components. This can be seen by constructing the auxiliary set $X_2$, and applying Lemma 2.
 More generally, we can translate any of the components of $X_3$ in a direction perpendicular to its principal 
phase and get a set with the same limit entropy. This shows that $X_3$ and $X_4$ have the same limit 
entropy. The construction in (c) involves a set formed by two parallel rectangular components, 
the upper component shown with dashed lines and the lower one coinciding with the lower component of $X_3$,
 and two wedge sets shown in dotted lines. The
 application of SSB leads to (\ref{apa1}). The construction in (d) involves the set $X_4$, two
 wedge sets and a flat rectangular set shown with dashed lines. Application of SSA leads to (\ref{apa2}).}
\label{multic}
\end{figure}

Now consider the construction of fig.(\ref{multic}) involving $d-n+1$ dimensional polyhedra. 
The set $X_1$ is 
formed by the union of two copies of a flat rectangular polyhedron $U=R\times[a]$
having principal planes that 
form a small angle $\theta$. Using the Lemma 2 
 we have that $\hat{\gamma}^{n-1}_S (X_2)\rightarrow \hat{\gamma}^{n-1}_S(X_1)+
 2s\, \text{vol}(R)\, \delta^n_1$ in the limit of small $a$ (see fig.
(\ref{multic})b).
By the same reason $\hat{\gamma}^{n-1}_S(X_2)\rightarrow \hat{\gamma}^{n-1}_S(X_3)+2s\, \text{vol}(R)
\, \delta^n_1$ in 
that limit 
(see fig.(\ref{multic})c). Then
 it follows that $\lim_{a\rightarrow 0} 
\hat{\gamma}^{n-1}_S(X_1)=\lim_{a\rightarrow 0} \hat{\gamma}^{n-1}_S(X_3)$. This means
 that we can move each of the components that are copies of $U$ in $X_1$ a direction perpendicular
 to their principal plane without changing the limit of the entropy for $a\rightarrow 0$. 
Thus, it also follows that 
$\lim_{a\rightarrow 0} \hat{\gamma}^{n-1}_S(X_4)=\lim_{a\rightarrow 0} \hat{\gamma}^{n-1}_S(X_3)$. 

Now we show that these two last limits have different expressions in terms of the function 
$\hat{\alpha}^n_S$ when $\theta \rightarrow 0$.  With the help of the auxiliary 
  sets drawn with dashed and dotted lines in the fig(\ref{multic})c and (\ref{multic})d,
and applying to these constructions
 SSB and SSA respectively we obtain 
\begin{eqnarray}
\lim_{a\rightarrow 0} \hat{\gamma}^{n-1}_S(X_3)\le \hat{\alpha}^n_S(a_1,...,a_{d-n})+2s \,\delta^n_1 \text{vol}(R)\,,
 \label{apa1} \\
\lim_{a\rightarrow 0} \hat{\gamma}^{n-1}_S(X_4)\ge \hat{\alpha}^n_S(2\,a_1,a_2,...,a_{d-n})+2s\, 
\delta^n_1 \text{vol}(R)\,,
\label{apa2}
\end{eqnarray}
where we have written explicitly the side lengths of the rectangular polyhedron that enter in the
arguments of $\hat{\alpha}^n_S$ and $(a_1,...,a_{d-n})$ are the side lengths of $R$. 
 Therefore $\hat{\alpha}^n_S(2\,a_1,a_2,...,a_{d-n})=\hat{\alpha}^n_S(a_1,a_2,...,a_{d-n})$, and 
from here 
\begin{equation}
\hat{\alpha}^n_S(a_1,...,a_{d-n})=\hat{\alpha}^n_S(a_1/2^r,...,a_{d-n})\,,\label{leftha}
\end{equation}
 for any
integer $r$. In the limit of large $r$ the right hand side of (\ref{leftha}) is just
$\hat{\alpha}^{n+1}_S(a_2,...,a_{d-n})$, which is zero by the induction hypothesis.
As the strongly subadditive an positive function $\hat{\alpha}^n_S$ is zero for all rectangular polyhedra 
it must be zero for
 all polyhedra, since they can be written as union of simplexes, which in turn can be
written as intersections of rectangular polyhedra. This completes the proof of the

\vspace{.3cm}
{\bf Proposition 2:} Let $S$ be a $d$ dimensional relativistic entropy function and 
$\alpha(X)=\gamma(X)-2s\,\text{vol(X)}$ the
 $d-1$ dimensional Euclidean function defined in Section IV. If $\alpha_0=0$ then 
$\alpha\equiv 0$.

\subsubsection{Proof of theorem 3}

Finally, we prove the theorem 3. We take $S$ with $\alpha_0=0$ and prove $S(X)$ is proportional 
to the spatial border area of $X$. From proposition 2 it follows for the flat polyhedra 
that $S(X\times [a])\rightarrow 2s\, \text{vol}(X)$ for any $d-1$ dimensional $X$ in the limit 
$a\rightarrow 0$. 
Thus, in this case the theorem 3 holds, since the boundary area of $X \times [a]$
 is $2\,\text{vol}(X)$.

\begin{figure}
\centering
\leavevmode
\epsfysize=4cm
\epsfbox{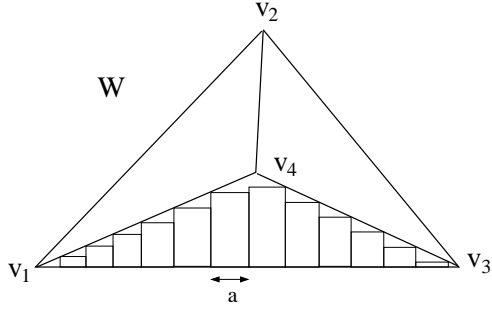}
\bigskip
\caption{A three-dimensional flat simplex $W$ with vertices $v_1$,$v_2$,$v_3$ and $v_4$.
 The picture shows  the projection on the plane of $v_1,v_2,v_2$. The limit taken in the text
 is when $v_4$ approaches a point $u$ on this plane. The decomposition shown for the lower simplex 
of vertices $v_1,v_3,v_4,u$ is in terms
 of wedge sets of width $a$, plus some small sets that can be formed by intersections of wedge and rectangular
 flat sets. Thus, in the limit $a\rightarrow 0$ and $v_4\rightarrow u$ we have the inequality (\ref{ooo}) 
from
 subadditivity.}
\label{sim1}
\end{figure}

A $d$ dimensional simplex W can be written as
$W=\{(x_1,...,x_d)/ (x_1,...,x_d)=r_1 v_1+...
+r_{d+1}v_{d+1},\,\, \text{with}\, \, r_i\ge 0 \,\,\text{for}\,\, i=1,...,d+1 \, \,\text{and}\,\, 
r_1+...+r_{d+1}=1\}$, where $v_1,...,v_{d+1}$ are the vectors pointing to the simplex vertices.
We are interested in a simplex in which $v_{d+1}=u+w$, where the vector $u$ lies inside the simplex 
phase $F$ with vertices $v_1,...,v_d$ and $w$ is a vector perpendicular to this phase. 
We are 
going to find an upper bound for the entropy $S(W)$ in the limit of small $\vert w\vert$. 
First decompose $W$ as union of the $d$ simplexes formed with
vertices $u$ and $v_{d+1}$, plus
any subset of $d-1$ vectors among the vectors $v_1,...,v_d$. 
We decompose each of these simplexes as union of wedge sets
with small
width $a$,
plus a number of small sets, as shown in the fig.(\ref{sim1}). In the
limit $\vert w\vert \rightarrow 0$
 the previous results show that the entropy of the wedge sets is proportional to 
the $d-1$ dimensional volume of their big phase.
 The entropy of the smaller sets is bounded by some constant times
 the $d-1$ volume of their projection on the plane of the simplex
 $F$, since these
sets can be formed by intersections of flat wedge and 
flat rectangular polyhedra.

Taking the limit of small $\vert w\vert$ and $a$ we then have
\begin{equation}
\lim_{\vert w\vert\rightarrow 0} S(W)\le 2s\,\text{vol}(F)\,.\label{ooo}
\end{equation}

\begin{figure}
\centering
\leavevmode
\epsfysize=5cm
\epsfbox{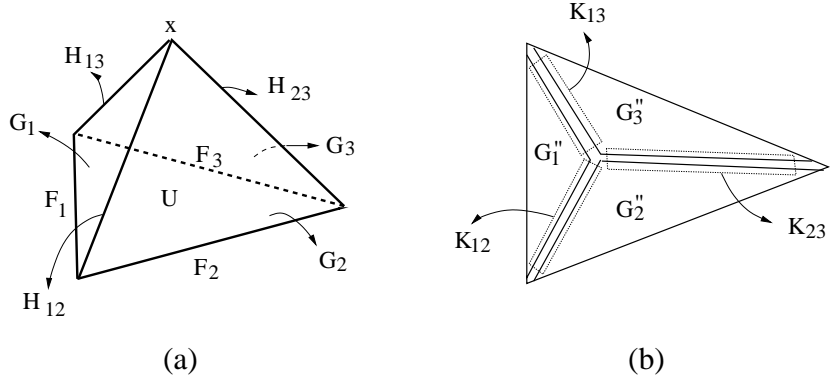}
\bigskip
\caption{(a)- The three dimensional set formed by the future 
domain of dependence of a two dimensional simplex $U$. The sets  $F_1$, $F_2$ and $F_3$ are the faces of $U$,
 and the point $x$ is at null distance from these faces.
$G_1$, $G_2$, and $G_3$ are the faces of the null boundary of the future domain of dependence,
 and $H_{ij}$ is the common phase between $G_i$ and $G_j$. (b)- A view of the set in (a) from the top.
 The simplexes $G_i^{\prime\prime}$ approach the null surfaces $G_i$ and thus they are flat simplexes.
 With the simplexes $G_i^{\prime\prime}$ and the commuting rectangular sets $K_{ij}$ plus some small
 sets of negligible entropy limit, we form a Cauchy surface for $U$. }
\label{sim2}
\end{figure}

Consider now any simplex $U$ with vertices  $u_1,...,u_{d+1}$ on the
hyperplane of time
$t=0$, and let $F_1,...,F_{d+1}$ be their faces.
Call $v$ the centre of the inscribed sphere,
the unique sphere that is
tangent to all the faces of $U$, and let $r$ be its radius. The point
 $x$ 
with spatial component $v$ at time $t=r$ is at null distance
to all $F_1,...,F_{d+1}$. Call $G_i$, with $i=1,...,d+1$
the null simplex determined by the phase $F_i$ and $x$, and
$H_{ij}$, with $i\neq j$, to the $d-1$ dimensional simplex that is
the common phase of $G_i$ and $G_j$ (see fig.(\ref{sim2})a).
A point $y$ with spatial component
$v$ at time
$t=r-\epsilon$ would
be at small distance from the faces of $U$ for $\epsilon$ small.
 The point
$y$ and a phase $F_i$ of $U$ determine a flat simplex $G_i^\prime$
in the above
sense, which converge to $G_i$ in the limit $\epsilon\rightarrow 0$.
The union of
the simplexes $G_i^\prime$ for $i=1,...,d+1$ form a
Cauchy surface for $U$.
We can slightly contract each of the simplexes $G_i^\prime$ in their
own hyperplanes to form the
simplexes
 $G_i^{\prime\prime}$, which  are
are at positive distance $a$ from each other (see fig.(\ref{sim2})b). 
 We can further impose to the faces of $G_i^{\prime\prime}$ and
 $G_j^{\prime\prime}$
 that are adjacent to each other to be parallel.
Then, we add flat elements $K_{ij}$ formed by the Cartesian
 product of simplexes $H_{ij}^\prime$,
slightly
smaller than the
$H_{ij}$, 
and a small interval $a+\epsilon^\prime$, in such a way to fill
 in the gaps between the different $G_i^{\prime\prime}$ and $G_j^{\prime\prime}$.
  We can take all the $K_{ij}$ and the $G_i^{\prime\prime}$
  to be commuting and the intersections of the $K_{ij}$ with
  $G_i^{\prime\prime}$ and $G_j^{\prime\prime}$ to be non empty
  sets of the form $H_{ij}^\prime$ times small intervals.
The union of all the $G_i^{\prime\prime}$, the $K_{ij}$,
 and a number of sets that are small in more than one perpendicular
 direction simultaneously and have vanishing small entropy when taking limits,
form a Cauchy surface for $U$.
In the limits $y\rightarrow x$ and $a\rightarrow 0$ the entropy for the flat
simplexes $G_i^{\prime\prime}$ becomes $2s$ times
 their big phase volume, which coincides with $\text{vol}(F_i)$.
 The entropy for $K_{ij}$ as well as the one for its intersection
 with $G_i^{\prime\prime}$ or $G_j^{\prime\prime}$ converges to
 $2s\, \text{vol}(H_{ij})$ in this limit (taking also $H_{ij}^\prime\rightarrow H_{ij}$ and 
$\epsilon^\prime\rightarrow 0$).
 Thus, the application of SSA to this geometrical construction leads
 to the inequality
 \begin{equation}
 S(U)\le 2s\,\left(\sum_{i=1}^{d+1} \text{vol}(F_i) -\sum_{i=2}^{d+1}
 \sum_{j=1}^{i-1} \text{vol}(H_{ij})\right)\,.
 \end{equation}
Now, as the hypersurfaces $G_i$ are null and have zero expansion, we have
 that the volume of $F_i$ is equal to the sum of the volumes of the
 $H_{ij}$ for all $j$ different from $i$.  Thus,
 \begin{equation}
 \sum_{i=1}^{d+1} \text{vol}(F_i) =2\sum_{i=2}^{d+1}
 \sum_{j=1}^{i-1} \text{vol}(H_{ij})\,.
\end{equation}
From here and the fact that $\sum_{i=1}^{d+1} \text{vol}(F_i)=
\text{area}(U)$ by definition, we have
\begin{equation}
S(U)\le s\,\text{area}(U)\,.\label{quipus}
\end{equation}

The formula (\ref{quipus}) allows us to obtain the corresponding
inequality for
any relativistic polyhedron $X$. This can be decomposed into simplexes
$U_i$ having empty
intersection or intersecting in adjacent faces $H_{ij}$.
We use the same procedure
as above. Slightly
contract the simplexes $U_i$ and fill the gaps in the Cauchy surface
 for $X$ with sets $K_{ij}$ formed by
product of simplexes $H_{ij}^\prime$ (converging to $H_{ij}$ when taking the
 limit) and small intervals. The $K_{ij}$ are taken to intersect with
  the $U_i$ and $U_j$ in sets that are the product of $H_{ij}^\prime$ and
   small intervals. Then, in the application of SSA to the construction,
 all the internal faces determined by adjacent simplexes $U_i$ and $U_j$ do not contribute. This
  is because, according to (\ref{quipus}), the contributions from the entropies of $U_i$
  and $U_j$ to the term corresponding to the phase $H_{ij}$ add to
  $2s\, \text{vol}(H_{ij})$, the contribution of $K_{ij}$ converges
   to $2s \, \text{vol}(H_{ij})$, while on the other side of the inequality,
    the intersection of $K_{ij}$ with $U_i$ and $U_j$ gives $4s\,
    \text{vol}(H_{ij})$.
Therefore, only the external faces of the $U_i$ (those that are also faces of $X$)
 contribute to the inequality, and we have
\begin{equation}
S(X)\le s\,\text{area}(X)\,\label{aaa}
\end{equation} 
 for all $X$.

Now take a rectangular polyhedron $R$ of sides $(a_1,...,a_d)$, and call its faces
$P_i$ with $i=1,..., d$ ($R$ has two faces equal to each $P_i$).
 Consider also the sets $P_i^\prime=P_i\times [3a]$ and
$P_i^{\prime\prime}=P_i\times [a,a,a]$, where $[a,a,a]$ is a one dimensional set 
of two components of size $a$ separated by a distance $a$. We have 
$\lim_{a\rightarrow 0} S(P_i^\prime)=2s\, \text{vol}(P_i)$ and
 $\lim_{a\rightarrow 0} S(P_i^{\prime\prime})=4s\, \text{vol}(P_i)$.
 We can form a big rectangular polyhedron $U$ by union of translated and non intersecting 
copies of $R$, the 
adjacent ones separated by a distance $a$, and filling the gaps by copies of the 
$P_i^\prime$ that intersect with the copies of $R$ in sets that are translated copies of 
$P_i^{\prime\prime}$. To complete $U$ we have to add sets that are small in more than one direction.
 Taking into account (\ref{aaa}) the limit of $S(U)/\text{vol}(U)$ for $U$ going to infinity is zero.  
Thus, the application of SSA to the construction leads in the limit of small $a$ and infinite $U$ to
\begin{equation}
S(R)\ge s\, \text{area}(R)\,.
\end{equation}
Thus, the equality $S(R)=s\, \text{area}(R)$ holds for any rectangular
polyhedron.

Any relativistic polyhedron $X$ is included into some rectangular polyhedron
$R$. Complete a Cauchy surface for 
$X$ with simplexes to form a Cauchy surface for $R$. Then,
 repeating the above procedure for dealing
 with the internal adjacent boundaries we obtain from SSA
\begin{equation}
S(X)\ge s\, \text{area}(X)\,.
\end{equation}
This finishes the proof of theorem 3.

\section{Discussion and conclusions}

 The properties of the quantum entropy make possible to deal with the
 geometric entropy in a geometrical way. However, even if it is believed
  that this quantity has a deep physical significance, it remains an abstract concept,
 wanting for a solid mathematical basis to compute it. In this sense, the principles employed in the paper
   to obtain the area law are very basic and general, and one can 
   expect them to be still available once a finite geometric entropy 
    is defined.

  A more immediate meaning can also be given to our results.
  They
  indicate that the entropy must be proportional to the area
  in the limit of infinite cutoff for any regularization scheme
  in QFT where covariant results are obtained asymptotically.  In other
  words, the leading divergence in the entropy must be proportional to
  the area for any set. We remark that the theorem shown in this paper
  tells that no subleading term exists in the continuum, excepting the
  constant. 

 This is indeed what all previous QFT calculations have shown for
 particular cases. Besides, we have also checked the area law
  for an spherical corona using the numerical method of ref. \cite{sred}. This case 
has also been studied in \cite{b1} in a related context. 
 This is a very interesting example since it allows to test 
 a very important and counterintuitive characteristic of the area solution,
   namely, that it
   is not ordered by inclusion outside the class of convex sets. As expected,
    the leading term in the entropy grows with the sum
    (with equal coefficients) of the areas of the internal and external
    surfaces of the corona.

There is a curiosity about the Euclidean solutions which could be related both 
 with the uniqueness of the relativistic entropy function and to the divergences. 
In the Euclidean case it is possible to construct many  entropy functions 
  averaging over translations and rotations positive and 
 strongly subadditive functions which are not Euclidean symmetric. 
 Among all these,   
we can find in particular some very interesting
 monotonic solutions that are exactly proportional to the area for the convex sets 
and less than the area for the non convex ones,  and where the entropy of two distant 
objects 'interact' by their mutual shadow. 
 We suspect that a classification of Euclidean
 solutions could be possible using this averaging method. 
Coming back to the relativistic case we see that a similar construction does not work  since the 
Lorentz group is non compact, leading to divergent integrals. 
 This is so even taking compact sets to make finite the integration over translations. 
 
 An interesting question related to the QFT calculations is the physical 
meaning of the coefficient relating area and entropy  and  the constant 
term. The coefficient of the area term was related to the Newton constant in \cite{w1,su} 
(see however \cite{ka1,mh,ot}), 
making contact with the black hole entropy. For $1+1$ dimensional conformal field theories  
it was shown in \cite{con} that the most divergent term for the single interval entropy is a 
constant (independent of size) proportional to the central charge. 
This coefficient corresponds to the combination $\gamma=\alpha+\beta$ according to (\ref{euler}). 
A natural question here is if a new parameter shows up in the expression for the multicomponent 
set entropy \cite{mhh}. In other words, the entropy can be simply proportional to the number 
of components ($\alpha=0$), 
or it can contain an additional constant term $\alpha$, with the same degree of divergence
 as $\beta$. It is possible that this last term is always subleading and negligible  in the continuum. This 
 is specially expected in more dimensions  
since the constant term requires a dimensionless parameter, 
 while the dimensionfull one in front of the area term diverges with the cutoff.    
Thus, the constant may represent a spurious solution
for $d>1$.
 
The entropy per unit area $s$ is a function 
$s_\rho$ of the Poincare symmetric state $\rho$. The combination of the inequalities (\ref{sll}) and 
Theorem 3 imply that given any two such states we have 
\begin{equation}
s_{\lambda_1 \rho_1+\lambda_2 \rho_2}=\lambda_1 s_{\rho_1}+\lambda_2 s_{\rho_2} \,, \label{fg}
\end{equation} 
 for positive $\lambda_1$ and  $\lambda_2$ with $\lambda_1+\lambda_2=1$. Thus the entropy per unit 
area is afin under mixing of states, a property known for the mean entropy per unit volume 
of traslational invariant states.  The physical meaning of eq. (\ref{fg}) is however unclear to the author. 

In a series of recent papers a very interesting phenomena has been pointed out \cite{b11,bb}. The authors
discovered that the entropy in a volume is not the only quantity that scales with the area of the 
given region but the same behavior is expected for a whole class of operators averaged using the 
 local density matrices. In particular they showed the area law for the energy fluctuations.  
 The results of this paper rely heavily on the use of the strong subadditive 
property of the entropy, and thus they can not be extended directly to operator correlations. However,
 it is possible that the method here could be generalized by introducing chemical potentials 
for suitable operators attached to the local regions. 

The proof in this paper uses explicitly arbitrarily small and large distances. 
The infiniteness of the space is specially relevant in the argument around the constructions of 
fig.(\ref{multic}), which involve large translations, and lead to the equation $\gamma(X)=\beta(X)$.
This is what excludes the possibility of finding a solution ordered by inclusion. 
However, it is not excluded that different type of solutions exist in de Sitter space. 
 This has compact Cauchy surfaces, and it is also
 maximally symmetric, what most probably can allow us to make 
a complete classification of solutions as the one given 
here for the Minkowski space \cite{mhh}.

The entropy of the black hole and the geometric entropy are 'empty' space
 quantities. There are a number of proposals for 
 general  bounds to the entropy of bounded systems which are suggested by the physics of black holes. 
They  have  been successfully checked mainly in the classical regime.  
The definition of a bounded system  
 requires a partition of the Hilbert space into a tensor product, and 
it is necessary to add boundary conditions. Thus, at the semiclassical level, the entropy bounds  
 should include a term coming from the vacuum correlations, 
even in empty space (see similar ideas in \cite{ss}). In principle, this could greatly tighten the inequalities.  
How do these bounds compare with our result in flat space?. 
Consider the covariant bound proposed in \cite{wa1}, which is related both with the Bekenstein bound \cite{b2} 
 and the generalized second law. Take a $d-1$ dimensional spatial surface $\Omega_1$
  and a null congruence ${\cal W}$ orthogonal to it, having non positive expansion, and 
 ending in another $d-1$ dimensional spatial surface $\Omega_2$. The bound states that the entropy 
crossing ${\cal W}$ 
 is bounded above by a constant times $\text{vol}(\Omega_1)-\text{vol}(\Omega_2)$. 
 Thus, this gives
 zero entropy for our polyhedra, since they have Cauchy surfaces formed by union of null flat surfaces
  having zero expansion. 

This touches upon the question of the geometric entropy for a more general type of subsets
 than the one considered here. The area and the constant term are not the most general solutions  
of the SSA inequalities for larger classes of subsets than the polyhedra. 
 Take for example the class of subsets with piecewise differentiable spatial corner. 
It is also a closed class under intersection and union of commuting sets. 
 As can be easily seen from fig(\ref{rela1})b, the integral over the spatial corner of any geometrical quantity 
that is positive and local gives place to a relativistic entropy function. Of course, 
the theorem 3 implies that these solutions are proportional to the area on the polyhedra.  As an example of such a 
local
 quantity  we can take the square of the expansion 
of the future null boundary of the given causally closed set.  
 In this case the entropy is non zero for the sphere and zero for every polyhedron approaching it. 
This shows that this type of solutions has a very pathological discontinuous behavior.

Excluding these solutions, probably unphysical,
 it seems we are left with two possibilities,  
either the covariant bound is wrong outside the 
classical domain, or the assumptions of this paper fail somewhere. However, it is also possible 
 and likely that the covariant bound has to be applied to a regularized notion of entropy  
 which is zero on Minkowski space. This is appealing, since the calculation
 of the black hole entropy using the  Euclidean path integral  
 requires the subtraction of the corresponding flat space quantity \cite{gh}. 

\section{Acknowledgments}
I thank the Abdus Salam International Centre for Theoretical Physics (ICTP) for 
 financial support and for the invitation to visit the institute where this work has been done.

\end{document}